\documentclass[%
 reprint,
 superscriptaddress,
 amsmath,amssymb,
 aps,
]{revtex4-2}

\usepackage{graphicx}
\usepackage{dcolumn}
\usepackage{bm}
\usepackage[mathlines]{lineno}

\usepackage{multirow}
\usepackage{bm}
\usepackage{textcomp}
\usepackage{longtable}
\usepackage{booktabs}
\usepackage{epsfig}
\usepackage{rotating}
\usepackage{xfrac}
\usepackage{threeparttable}
\usepackage{lipsum} 
\usepackage{float}
\RequirePackage{hyperref}
\hypersetup
{
unicode=false,%
pdftoolbar=true,
pdfmenubar=true,
pdffitwindow=false,
pdfstartview=FitH,
pdftitle={PRC},
pdfauthor={LAM Yi Hua},
pdfsubject={Nuclear Astrophysics},
pdfcreator={LAM Yi Hua},
pdfproducer={LAM Yi Hua},
pdfkeywords={NucAstro},
bookmarksnumbered,%
colorlinks=true,%
linkcolor=magenta,%
citecolor=blue,
filecolor=cyan,%
urlcolor=blue,
breaklinks, 
plainpages=false}
\usepackage{CJKutf8}
\newcommand{\cntextSimKai}[1]{\begin{CJK*}{UTF8}{gkai}#1\end{CJK*}}
\newcommand{\cntextTraKai}[1]{\begin{CJK*}{UTF8}{bkai}#1\end{CJK*}}
\newcommand{\jptextJap}[1]{\begin{CJK*}{UTF8}{min}#1\end{CJK*}}
\usepackage{xcolor}
\definecolor{Mycolor}{HTML}{C5E0B3}
\definecolor{red}{HTML}{ee6677}
\definecolor{blue}{HTML}{4477aa}
\definecolor{teal}{HTML}{009988}
\definecolor{green}{HTML}{228833}
\definecolor{magenta}{HTML}{ee3377}
\definecolor{cyan}{HTML}{66ccee}
\definecolor{yellow}{HTML}{ccbb44}
\definecolor{grey}{HTML}{bbbbbb}
\begin{document}

\preprint{APS/123-QED}

\title{New $^{63}$Ga(p,$\gamma$)$^{64}$Ge and $^{64}$Ge(p,$\gamma$)$^{65}$As reaction rates \\ corresponding to the temperature regime of thermonuclear x-ray bursts}

\author{Ning~Lu~(\cntextTraKai{盧寧})}
\affiliation{CAS Key Laboratory of High Precision Nuclear Spectroscopy, \href{https://ror.org/03x8rhq63}{Institute of Modern Physics}, Chinese Academy of Sciences, Lanzhou 730000, China}%
\affiliation{School of Nuclear Science and Technology, \href{https://ror.org/01mkqqe32}{Lanzhou University}, Lanzhou 730000, China}
\affiliation{School of Nuclear Science and Technology, \href{https://ror.org/05qbk4x57}{University of Chinese Academy of Sciences}, Beijing 100049, China}

\author{Yi~Hua~Lam~(\cntextTraKai{藍乙華})}%
 \email{lamyihua@impcas.ac.cn}
 \thanks{Present address: Zhejiang Key Laboratory of Quantum State Control and Optical Field Manipulation, Department of Physics, \href{https://ror.org/03893we55}{Zhejiang Sci-Tech University}, 310018 Hangzhou, China}
\affiliation{CAS Key Laboratory of High Precision Nuclear Spectroscopy, \href{https://ror.org/03x8rhq63}{Institute of Modern Physics}, Chinese Academy of Sciences, Lanzhou 730000, China}%
\affiliation{School of Nuclear Science and Technology, \href{https://ror.org/05qbk4x57}{University of Chinese Academy of Sciences}, Beijing 100049, China}
\affiliation{Astrophysical Big Bang Laboratory (ABBL), Cluster for Pioneering Research, \href{https://ror.org/01sjwvz98}{RIKEN}, Wako, Saitama 351-0198, Japan}

\author{Alexander~Heger}
\affiliation{School of Physics and Astronomy, \href{https://ror.org/02bfwt286}{Monash University}, Vic 3800, Australia}%
\affiliation{OzGrav-Monash -- Monash Centre for Astrophysics, School of Physics and Astronomy, \href{https://ror.org/02bfwt286}{Monash University}, Vic 3800, Australia}%
\affiliation{Center of Excellence for Astrophysics in Three Dimensions (ASTRO-3D), School of Physics and Astronomy, \href{https://ror.org/02bfwt286}{Monash University}, Vic 3800, Australia}
\affiliation{Joint Institute for Nuclear Astrophysics, \href{https://ror.org/05hs6h993}{Michigan State University}, East Lansing, MI 48824, USA}

\author{Zi~Xin~Liu~(\cntextSimKai{刘子鑫})}
\affiliation{Advanced Energy Science and Technology, Guangdong Laboratory, Huizhou 516000, China}
\affiliation{CAS Key Laboratory of High Precision Nuclear Spectroscopy, \href{https://ror.org/03x8rhq63}{Institute of Modern Physics}, Chinese Academy of Sciences, Lanzhou 730000, China}

\author{Hidetoshi~Yamaguchi (\jptextJap{山口英斉})}
\affiliation{Center for Nuclear Study(CNS), \href{https://ror.org/057zh3y96}{the University of Tokyo}, RIKEN campus, 2-1 Hirosawa, Wako, Saitama 351-0198, Japan} 
\affiliation{\href{https://ror.org/052rrw050}{National Astronomical Observatory of Japan}, 2-21-1 Osawa, Mitaka, Tokyo 181-8588, Japan}



\date{\today}

\begin{abstract}
We compute the $^{63}$Ga(p,$\gamma$)$^{64}$Ge and $^{64}$Ge(p,$\gamma$)$^{65}$As thermonuclear reaction rates using the latest experimental input supplemented with theoretical nuclear spectroscopic information. The experimental input consists of the latest proton thresholds of $^{64}$Ge and $^{65}$As, and the nuclear spectroscopic information of $^{65}$As, whereas the theoretical nuclear spectroscopic information for $^{64}$Ge and $^{65}$As are deduced from the full \emph{pf}-shell space configuration-interaction shell-model calculations with the GXPF1A Hamiltonian. Both thermonuclear reaction rates are determined with known uncertainties at the energies that correspond to the Gamow windows of the temperature regime relevant to type I x-ray bursts, covering the typical temperature range of the thermonuclear runaway of the GS~1826$-$24 periodic bursts and SAX~J1808.4$-$3658 photospheric radius expansion bursts. 
\end{abstract}


\maketitle

\section{Introduction}
\label{sec:intro}
The range of synthesized nuclei and the impact on the nuclear energy generation during an onset of type I x-ray burst (XRB) are sensitive to the proton thresholds and thermonuclear reaction rates around some key branching points and waiting points, e.g., the $^{22}$Mg branching point \citep{Randhawa2020,Hu2021,Jayatissa2023} and the $^{56}$Ni \citep{Valverde2018,Valverde2019,Kahl2019,Lam2022b}, $^{60}$Zn, and $^{64}$Ge waiting points \citep{Wormer1994,Woosley2004,Parikh2008,Parikh2009,Cyburt2016,Lam2016}. Obtaining the important reaction rates surrounding the waiting points like $^{56}$Ni and $^{64}$Ge with (low and) known uncertainties \citep{Lam2022a,Lam2022b} permits us to further constrain the important astrophysical parameters for state-of-the-art XRB models and to enhance the capability of these models producing the key properties, i.e., burst light curves, peaks, fluences, recurrence times, Eddington fluxes, and persistent fluxes, closely matched with the observed counterparts \citep{Johnston2020}.

These key properties manifest the mutual influence between nuclear energy generation and thermo-hydrodymamics during the evolution of thermonuclear runaway at the accreted envelope of an accreting neutron star \citep{Woosley2004,Heger2007,Goodwin2019,Johnston2020,Lam2022a,Lam2022b}. Such essential mutual influence is taken into account by one-dimensional multi-zone hydrodynamic XRB models instantiated by, e.g., \textsc{Kepler} \citep{Woosley2004,Heger2007}, \textsc{Mesa} \citep{MESA2015}, \textsc{Agile} \citep{AGILE,Fisker2008}, and \textsc{Shiva} \citep{Jose2010}. For the one-zone model \citep{Schatz2001} and post-processing \citep{Parikh2008, Bojazi2014, NucNet, SkyNet} models, the important mutual influence is completely missing for calculating the evolution of the accreted envelope, however. Therefore, the snapshots of temperature-density profiles generated by the XRB models with hydrodynamics are needed for post-processing models, but the changes of nuclear energy generation due to any implemented new reaction rates do not mutually feedback with the fixed temperature-density profiles.

Having the high-fidelity XRB models with such capability allows us to probe the details of the rapid-proton-capture (rp-) process path, nucleosynthesis, neutron star surface gravity, distance, and gravitational redshift corresponding to the selected and studied XRB scenarios of the respective x-ray bursters. Furthermore, the highly-constrained astrophysical parameters are the essential ingredients for us to understand the dense matter of accreting neutron stars \citep{Dohi2020,Dohi2022,Xie2024}, thermo-hydrodynamics at the accreted envelope \citep{Woosley2004}, composition of the accreted material associated with the evolution of the companion star \citep{Heger2007,Johnston2020,Goodwin2019}, which forms the low-mass x-ray binary system together with the accreting neutron star. 

Both $^{63}$Ga(p,$\gamma$)$^{64}$Ge and $^{64}$Ge(p,$\gamma$)$^{65}$As thermonuclear reactions connect the ZnGa and GeAs cycles \citep{Lam2022a}, which include the $^{60}$Zn and $^{64}$Ge waiting points. The GeAs cycles are weakly formed due to the slow two-proton sequential capture on $^{64}$Ge as found by \citet{Lam2022a} prior to \citet{Zhou2023} with deducing and reassessing the $^{65}$As(p,$\gamma$)$^{66}$Se reaction rate. The proton threshold of $^{66}$Se was determined by using the experimental mirror nuclear masses, theoretical mirror displacement energies, and full $pf$-model space shell-model calculations. The self-consistent relativistic Hartree-Bogoliubov approach (see Refs.~\citep{Liu2023,Liu2024} for the details) with the explicit density-dependent meson-\textcolor{black}{exchange} (DD-ME2) effective interaction \citep{Lalazissis2005} was used to compute the mirror displacement energies. The theoretical proton threshold of $^{66}$Se, $S_\mathrm{p}$($^{66}$Se)~$\!=\!2.469\pm0.054$~MeV, agrees well with the precisely measured counterpart by \citet{Zhou2023}, $S_\mathrm{p}$($^{66}$Se)~$\!=\!2.465\pm0.074$~MeV, using the experimental cooler-storage ring (CSRe) \citep{CSRe} at the Heavy Ion Research Facility in Lanzhou (HIRFL) \citep{HIRFL}.

The connection between the ZnGa and GeAs cycles can be further constrained with reassessing both $^{63}$Ga(p,$\gamma$)$^{64}$Ge and $^{64}$Ge(p,$\gamma$)$^{65}$As reaction rates. 
Recently, the impact of several proton thresholds deduced from the newly measured proton-rich nuclear masses on the periodic burst light-curve profile of GS~1826$-$24 burster was studied by \citet{Zhou2023}. These proton-rich nuclei are $^{63}$Ge, $^{64,65}$As, and $^{66,67}$Se. Their study shows that these new proton thresholds from new masses cause the reduction of proton-captures at $^{64}$Ge and the pronounced effect on reproducing the periodic burst light-curve profile of GS~1826$-$24 clocked burster. 
However, their study was merely based on the updated proton thresholds without reassessing the relevant (p,$\gamma$) forward and reverse reaction rates, i.e., 
$^{63}$Ge(p,$\gamma$)$^{64}$As, 
$^{64}$Ge(p,$\gamma$)$^{65}$As, 
$^{65}$As(p,$\gamma$)$^{66}$Se, and
$^{66}$As(p,$\gamma$)$^{67}$Se. 
Moreover, the initial models used by them are not constrained to reproducing the observed burst peak, tail end, and fluence. 
In fact, the proper and essential procedure of examining the impact of new masses requires the reassessment of relevant thermonuclear reaction rates, e.g., the recent studies performed by \citet{Valverde2018}, and also the constrained XRB models on the observed burst peak, tail end, and fluence (see Refs.~\citep{Lam2022a, Lam2022b} for the details). 

In this work, we reassess the (p,$\gamma$) forward and reverse reaction rates of $^{63}$Ga(p,$\gamma$)$^{64}$Ge and $^{64}$Ge(p,$\gamma$)$^{65}$As using the latest experimental input supplemented with theoretical nuclear spectroscopic information. The experimental input consists of the latest proton thresholds of $^{64}$Ge \citep{Clark2007,Schury2007,AME2020} and $^{65}$As \citep{Zhou2023}, and the nuclear spectroscopic information of $^{65}$As \citep{Obertelli2011}, whereas the theoretical nuclear spectroscopic information is deduced from the full \emph{pf}-shell space configuration-interaction shell-model calculations with the GXPF1A Hamiltonian \citep{Honma2004,Honma2005}. Both thermonuclear reaction rates are determined at the energies that correspond to the Gamow windows of the temperature regime relevant to type I x-ray bursts, covering the typical temperature range of the thermonuclear runaway of the GS~1826$-$24 periodic bursts and SAX~J1808.4$-$3658 photospheric radius expansion bursts. The pioneer periodic-burst model of GS~1826$-$24 burster was constructed by \citet{Heger2007}, and the first sequential-photospheric-radius-expansion-burst model of SAX~J1808.4$-$3658 burster was deduced by \citet{Johnston2018}. The periodic bursts of GS~1826$-$24 burster are the most investigated by both astrophysics and nuclear astrophysics communities.


In this paper, we present the formalism for constructing the reaction rates in Section~\ref{sec:rates}. The new $^{63}$Ga(p,$\gamma$)$^{64}$Ge and $^{64}$Ge(p,$\gamma$)$^{65}$As reaction rates are discussed in detail in Section~\ref{sec:discussion}. The conclusion of this work is given in Section~\ref{sec:summary}. 

\section{THERMONUCLEAR REACTION RATE}
\label{sec:rates}

\subsection{Formalism}
\label{sec:rate_formalism}

The total thermonuclear proton capture reaction rate is given by the sum of reaction rates of resonant and direct proton captures. The total rate can be expressed as follows~\citep{Fowler1964,Rolfs1988}.
\begin{eqnarray}
\label{eq:total}
N_\mathrm{A}\langle \sigma v \rangle = && \sum_i(N_\mathrm{A}\langle \sigma v \rangle_\mathrm{res}^i+N_\mathrm{A}\langle \sigma v \rangle_\mathrm{DC}^i)\nonumber\\
&&\times\frac{(2J_i+1)e^{-E_i/kT}}{\sum_n(2J_n+1)e^{-E_n/kT}} \, ,
\end{eqnarray}
of which each contributing resonant (and direct proton) captures are folded with the respective individual population factors. The acronyms ``res'' and ``DC'' denote resonant proton capture and direct proton capture, respectively (see Refs.~\cite{Lam2016, Lam2022a, Lam2022b} for the detailed notations). 

\subsection{Resonant rates}
\label{sec:Resonant}
The isolated narrow resonant reaction rate for the proton capture on the target nucleus in its initial state $i$, $N_\mathrm{A}\langle \sigma v \rangle_\mathrm{res}^i$, can be calculated by a sum over all relevant compound nucleus states~$j$ above the proton separation energy, $S_\mathrm{p}$~\citep{Rolfs1988,Iliadis2007}. The resonant rate can be described as~\citep{Fowler1967,Schatz2005},
\begin{eqnarray}
\label{eq:res}
N_\mathrm{A}\langle \sigma v \rangle_\mathrm{res}^i = && 1.54 \times 10^{11} (\mu T_9)^{-3/2} \times \sum_j \omega\gamma_{ij}  \nonumber\\
                                && \times \;\; \mathrm{exp} \left (-\frac{11.605E^{ij}_\mathrm{res}}{T_9} \right)\, 
\end{eqnarray} 
where $E^{ij}_\mathrm{res}\!=\!E^j_\mathrm{x} - S_\mathrm{p} - E_i$ is the resonance energy in the center-of-mass system calculated from the energies of the initial $E_i$ and compound nucleus state $j$ with the energy $E^j_\mathrm{x}$, and $T_9$ is the temperature in Giga Kelvin (GK). Here we take into account the ground state (g.s.) and thermally excited states, $E_i$, in the target nucleus. $\mu$ is the reduced mass of the entrance channel in atomic mass units. 

The resonance strengths, $\omega\gamma_{ij}$, in Eq.~(\ref{eq:res}) are given in units of MeV and can be expressed as,

\begin{eqnarray}
\label{eq:resonantStrength}
\omega\gamma_{ij}=\frac{2J_j+1}{2(2J_i+1)}\frac{\Gamma_\mathrm{p}^{ij}\times\Gamma_\gamma^j}{\Gamma_\mathrm{total}^j} \, 
\end{eqnarray}
where $J_i$ is the target spin and $J_j$, $\Gamma_\mathrm{p}^{ij}$ , $\Gamma_\gamma^j$, and $\Gamma_\mathrm{total}^j$ are spin, proton-decay width, $\gamma$-decay width, and total width of the compound nucleus state~$j$, respectively. The total width is defined as $\Gamma_\mathrm{total}^j$=$\Gamma_\gamma^j+\Gamma_\mathrm{p}^{ij}$, assuming other channels are closed in the excitation energy range considered in this work. The proton width is expressed as $\Gamma_\mathrm{p}$ = $\sum_{\mathtt{nlj}} C^2S(\mathtt{nlj}) \Gamma_\mathrm{s.p.}(\mathtt{nlj})$, where $C^2 S(\mathtt{nlj})$ is the single-proton-transfer spectroscopic factor associated with the single-proton width\textcolor{black}{, $\Gamma_\mathrm{s.p.}$}. The \textcolor{black}{$\Gamma_\mathrm{s.p.}$} can be obtained from \textcolor{black}{$\Gamma_\mathrm{s.p.}$=$\frac{3\hbar^2}{\mu R^2}P_{\mathtt{l}}(E)$}~\citep{Wormer1994,Herndl1995}, \textcolor{black}{where $R\!=\!r_0\!\times\!(1\!+\!A_\mathrm{T})^{1/3}$~fm (with $r_0\!=\!1.25$~fm and $A_\mathrm{T}$ is the target mass number)} is the nuclear channel radius\textcolor{black}{, and $P_\mathtt{l}$ is the Coulomb barrier penetration factor. Alternatively, $\Gamma_\mathrm{s.p.}$ can be estimated from the} proton scattering cross section\textcolor{black}{s in a} Woods-Saxon potential well \textcolor{black}{by adjusting the potential depth to reproduce the known proton energies. The same procedure was implemented in Refs.~\citep{Lam2016,Lam2022a,Lam2022b} for obtaining the $\Gamma_\mathrm{s.p.}$ for the $^{64}$Ge(p,$\gamma$)$^{65}\!$As, $^{65}\!$As(p,$\gamma$)$^{66}$Se, and $^{57}$Cu(p,$\gamma$)$^{58}$Zn reaction rates.} 
The \textcolor{black}{$\Gamma_\mathrm{s.p.}$} yielded from these two methods agree well with each other, and the maximum difference is merely \textcolor{black}{up to} \textcolor{black}{$46.7$~\%} \textcolor{black}{for the present work}. We only take into account the $\gamma$-decay widths from the M1 and E2 electromagnetic transitions for the resonance states as their contribution are exponentially higher than the M3 and E4 transitions. 

The experimental nuclear structure information above the proton thresholds and within the respective Gamow energy for the $^{63}$Ga(p,$\gamma$)$^{64}$Ge and $^{64}$Ge(p,$\gamma$)$^{65}$As reactions is somehow scarce and limited. For instance, the experimentally determined structure information above the proton threshold of $^{64}$Ge is rather scarce, especially the excited states of angular momenta corresponding to the compound states of $^{63}$Ga+p. The case of $^{65}$As is even limited, only the ground state binding energy and the first excited state of $^{65}$As were recently measured to be \textcolor{black}{$-\!46.806\!\pm\!0.042    $}~MeV~\citep{Zhou2023} and \textcolor{black}{$E_\mathrm{x}\!=\!0.187\!\pm\!0.003$}~MeV~\citep{Obertelli2011}, respectively.

\begin{figure}[t!]%
\centering
\includegraphics[width=0.45\textwidth]{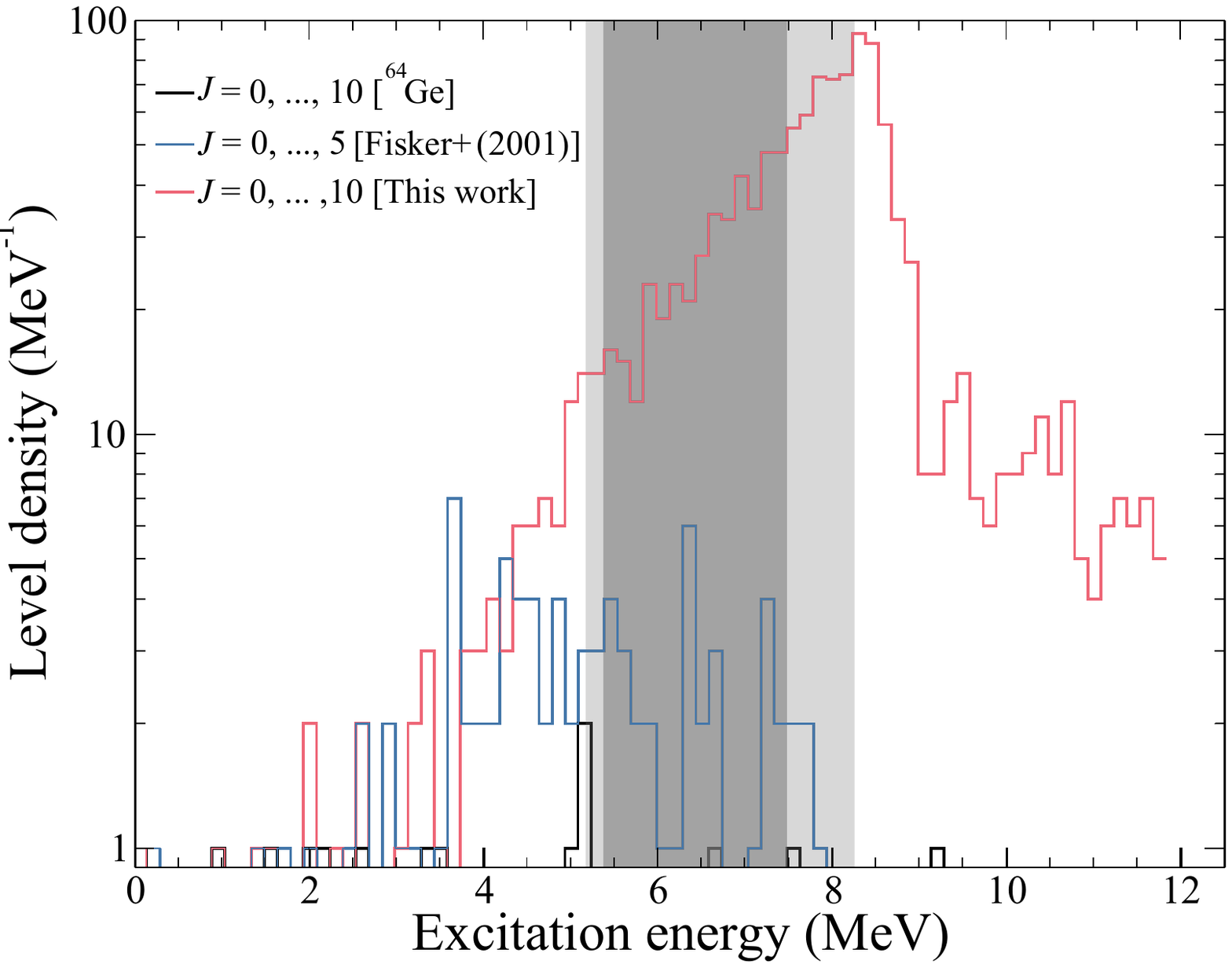}
\caption{The level density of $^{64}$Ge with energy bin of 150~keV. The level density of $^{64}$Ge determined from available experimental data 
(black line) \citep{NNDC_64Ge_2021}. 
The $^{64}$Ge level density deduced from the level scheme calculated by \citet{Fisker2001} using the truncated \emph{pf}-shell space shell-model calculations with the KBF interaction for angular momenta, $J^\pi\!=\!0^+$, ..., $5^+$ (blue line). The present full \emph{pf}-shell space shell-model calculations using the GXPF1A interaction for a set of angular momenta, $J^\pi\!=\!0^+$, ..., $10^+$ (red line). The laboratory Gamow energy (light gray zone) corresponds to the $^{63}$Ga(p,$\gamma$)$^{64}$Ge reaction rate at the temperature regime of $0.1\!\leq\!T\mathrm{(GK)}\!\leq\!2$, whereas the dark gray zone is the temperature regime sensitive to the GS~1826$-$24 clocked bursts and SAX~J1808.4$-$3658 PRE bursts.}
\label{fig:level_density_64Ge}
\end{figure}

In order to supplement the nuclear structure information, i.e., energy levels of $^{64}$Ge, spectroscopic factors, proton and gamma partial widths needed for deducing the $^{63}$Ga(p,$\gamma$)$^{64}$Ge reaction rate with identified uncertainties, we perform large scale shell-model calculations with the full \emph{pf}-shell space to obtain the properties of resonances at the Gamow energy sensitive to the XRB temperature range using the GXPF1A Hamiltonian~\citep{Honma2004,Honma2005}.

%

With the consideration of populating the resonance states at Gamow energy for constructing the $^{64}$Ge level density, we calculate $60$, $130$, $180$, $170$, $180$, $140$, $130$, $70$, and $70$ states for each spin-parity of $0^+$, $1^+$, $2^+$, $3^+$, $4^+$, $5^+$, $6^+$, $7^+$, and $8^+$, respectively. These states are calculated by diagonalizing the $^{64}$Ge nuclear Hamiltonian matrices of dimensions up to \textcolor{black}{$1.087\times10^{9}$}. The dimensions of nuclear Hamiltonian matrices of $^{63}$Ga are up to \textcolor{black}{$1.433\times10^{9}$}. These huge matrices are diagonalized using the \textsc{KShell} code \citep{KShell} with double precision. The calculations are performed on the \textcolor{black}{FDR5 cluster of Academia Sinica Grid Computing Center (ASGC) of Academia Sinica, Taiwan}.

We then deduce and compare theoretical and experimental level density associated to $^{64}$Ge in Fig.~\ref{fig:level_density_64Ge}. To compare with a sum of the experimental level density of $^{64}$Ge, 
we also calculate $70$ states for each spin-parity of $9^+$ and $10^+$. The energies of $0^+_{60}$, $1^+_{130}$, $2^+_{180}$, $3^+_{170}$, $4^+_{180}$, $5^+_{140}$, $6^+_{130}$, $7^+_{70}$, and $8^+_{70}$ are \textcolor{black}{8.687~MeV, 8.889~MeV, 8.605~MeV, 8.485~MeV, 8.556~MeV, 8.617~MeV, 8.955~MeV, 8.919~MeV, and 9.663~MeV}, respectively. The first $9^+$ and $10^+$ states are \textcolor{black}{6.923~MeV and 7.103~MeV}, respectively. The level density at the Gamow energy, \textcolor{black}{5.206-8.280~MeV} (light gray zone in Fig.~\ref{fig:level_density_64Ge}), is mainly populated by $1^+$, $2^+$, $3^+$, $4^+$, $5^+$, and $6^+$ states, with more than $70$ states for each angular momenta. These states are exhausted above \textcolor{black}{8.5~MeV} causing the drop of the theoretical level density along \textcolor{black}{$E_\mathrm{x}\!=\!8.28$-$12$~MeV} (red line in Fig.~\ref{fig:level_density_64Ge}). Such decrease of theoretical level density at $E_\mathrm{x}\!>\!8.28$~MeV does not affect the present construction of the $^{63}$Ga(p,$\gamma$)$^{64}$Ge reaction rate for the Gamow energy, \textcolor{black}{5.206-8.280~MeV} (light gray zone in Fig.~\ref{fig:level_density_64Ge}). The angular momenta of $0^+$, $1^+$, $2^+$, $3^+$, $4^+$, and $5^+$ had been taken into account by \citet{Fisker2001}, however, based on the truncated \emph{pf}-shell space calculation and the KBF interaction \citep{Poves2001}. The present theoretical estimation using the full \emph{pf}-shell space indicates that $^{64}$Ge has a high level density at the Gamow energy, whereas the level densities deduced from the \citet{Fisker2001} work (blue line in Fig.~\ref{fig:level_density_64Ge}) and scarce experimental data (black line in Fig.~\ref{fig:level_density_64Ge}) are about one order of magnitude lower than the present level density for the Gamow energy, especially the energy range, \textcolor{black}{5.450-7.538~MeV}, corresponding to the temperature region sensitive to the GS~1826$-$24 clocked bursts and SAX~J1808.4$-$3658 PRE bursts (dark gray zone in Fig.~\ref{fig:level_density_64Ge}). 

\begin{figure}[t]
\includegraphics[width=8.7cm, angle=0]{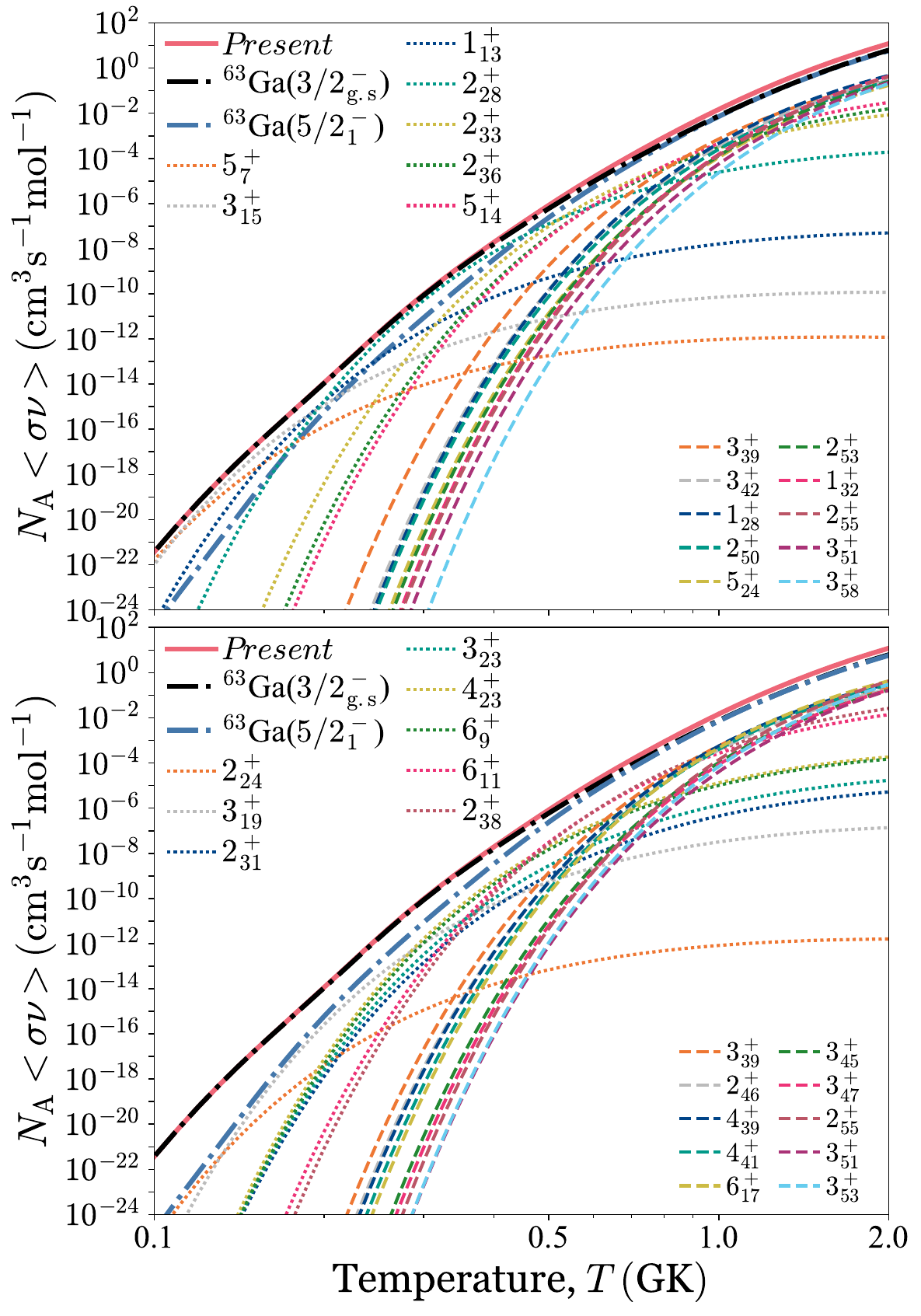}
\caption{\label{fig:rp_63Ga_64Ge_contri}{
The dominant resonances contributing to the $^{63}$Ga(p,$\gamma$)$^{64}$Ge thermonuclear reaction rates in the temperature region of XRB interest. 
Top panel: the main contributing resonances of proton captures on the $3/2^-_\mathrm{g.s.}$ state of $^{63}$Ga.
Bottom panel: the main contributing resonances of proton captures on the first excited state, $5/2^-_1$, of $^{63}$Ga. 
Both total contributing resonances of proton captures on $3/2^-_\mathrm{g.s.}$ and on $5/2^-_1$ states are indicated as black dash-dotted and blue dash-dotted lines, respectively. The total present rate (red line) is shown for comparison. See Table~\ref{tab:properties_64Ge}. 
}}
\end{figure}

\begin{figure}[t]
\includegraphics[width=8.7cm, angle=0]{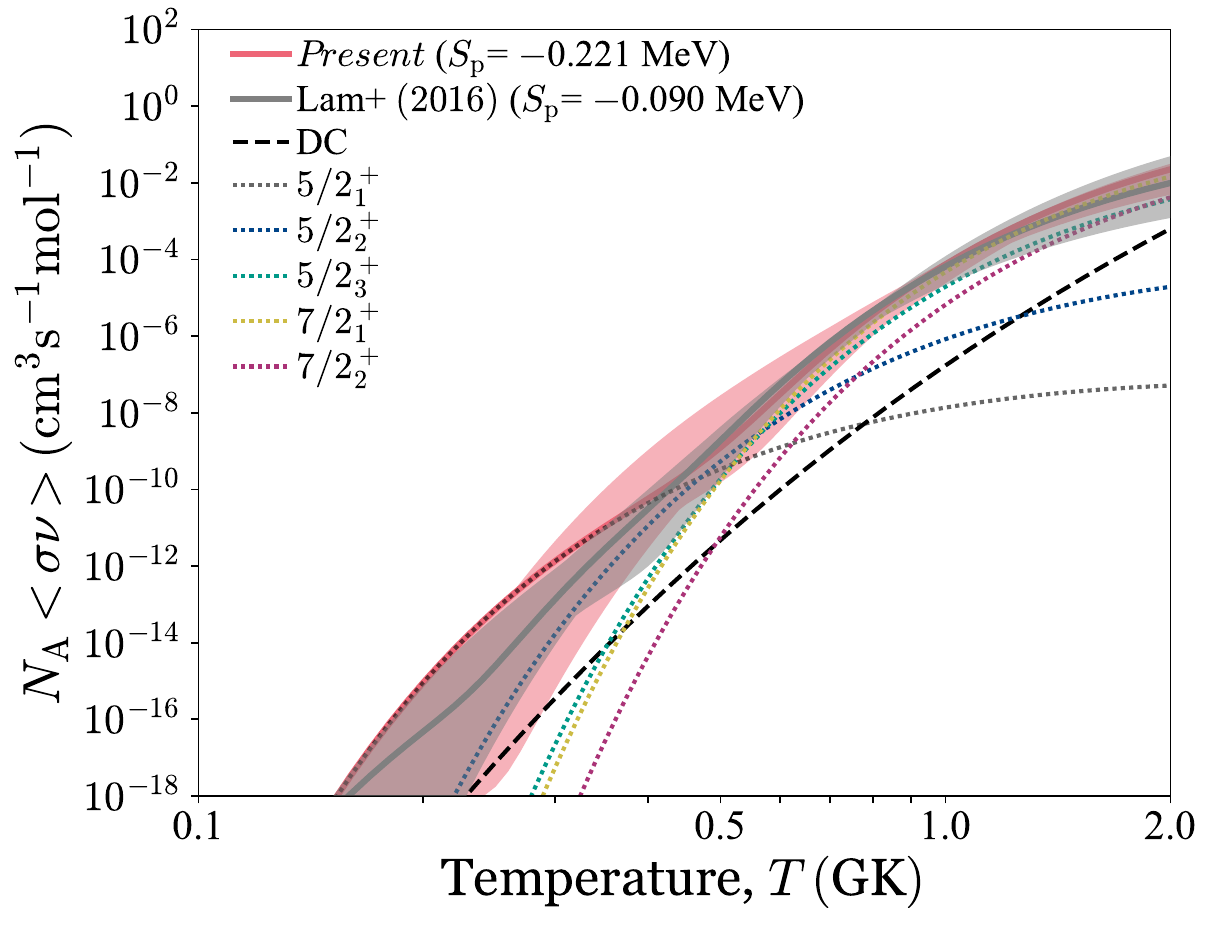}
\caption{\label{fig:rp_64Ge65As_contri}{
The dominant resonances contributing to the $^{64}$Ge(p,$\gamma$)$^{65}$As thermonuclear reaction rates in the temperature region of XRB interest. 
The main contributing resonances of proton captures on the $0^+_\mathrm{g.s.}$ state of $^{64}$Ge (dotted lines). The uncertainties of \citet{Lam2016} and the present rates are indicated as gray and red zones, respectively. See Table~\ref{tab:properties_65As}.
}}
\end{figure}


\begin{figure}[t]
\includegraphics[width=8.7cm,angle=0]{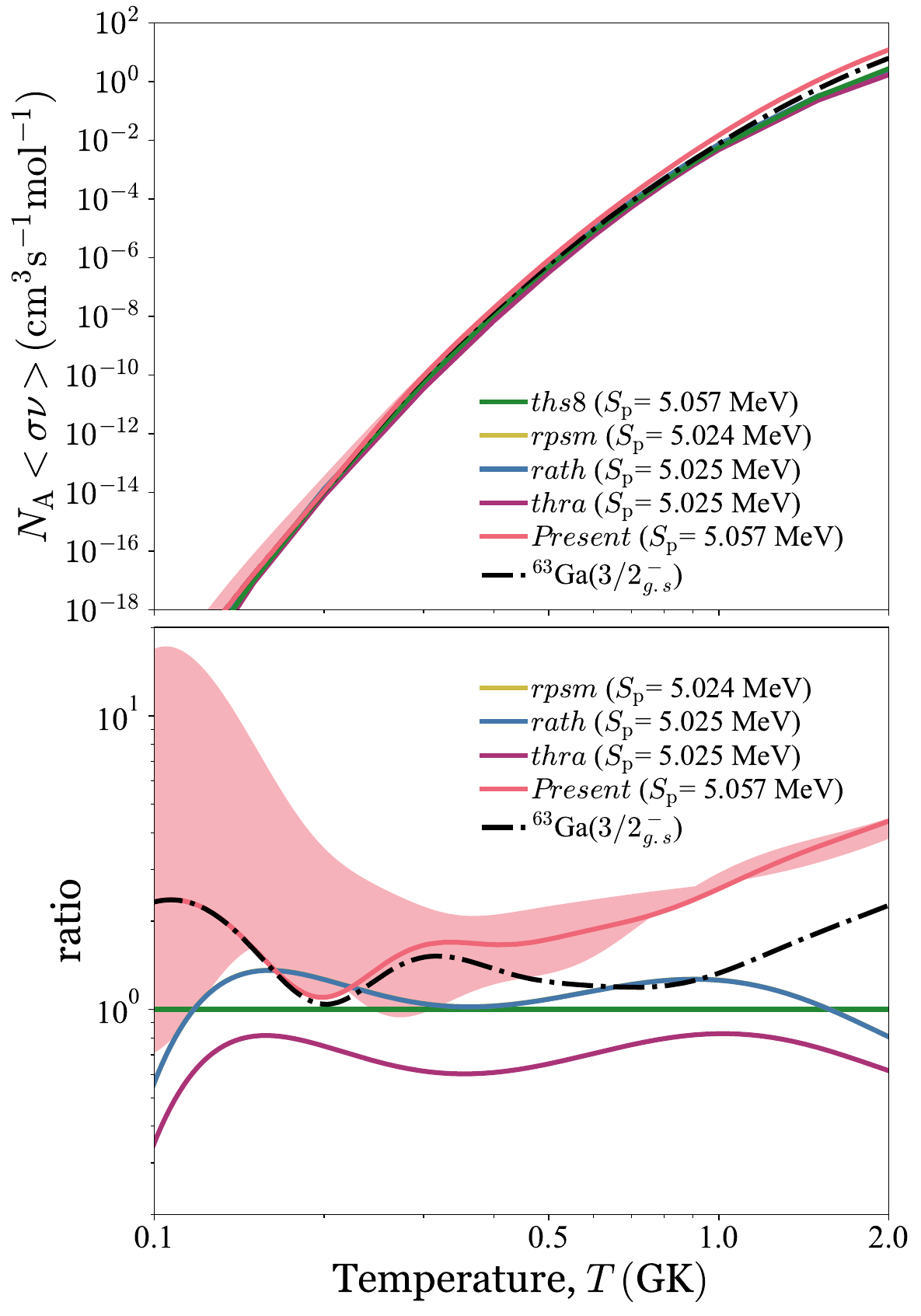}
\caption{\label{fig:rp_63Ga_64Ge}The comparison of $^{63}$Ga(p,$\gamma$)$^{64}$Ge thermonuclear reaction rates. Top Panel: \emph{ths}8, \emph{rath}, \emph{rpsm}, and \emph{thra} rates are the available rates compiled in JINA REACLIB v2.2 \citep{Cyburt2010} and \emph{ths}8 is the recommended rate published in part of the JINA REACLIB~v2.2 release. 
Bottom Panel: the respective ratio of the reaction rates presented in the top panel to the recommended \emph{ths}8 rate. The uncertainty of the present rate is indicated as red zone.
}
\end{figure}

\begin{figure}[t]
\includegraphics[width=8.7cm,angle=0]{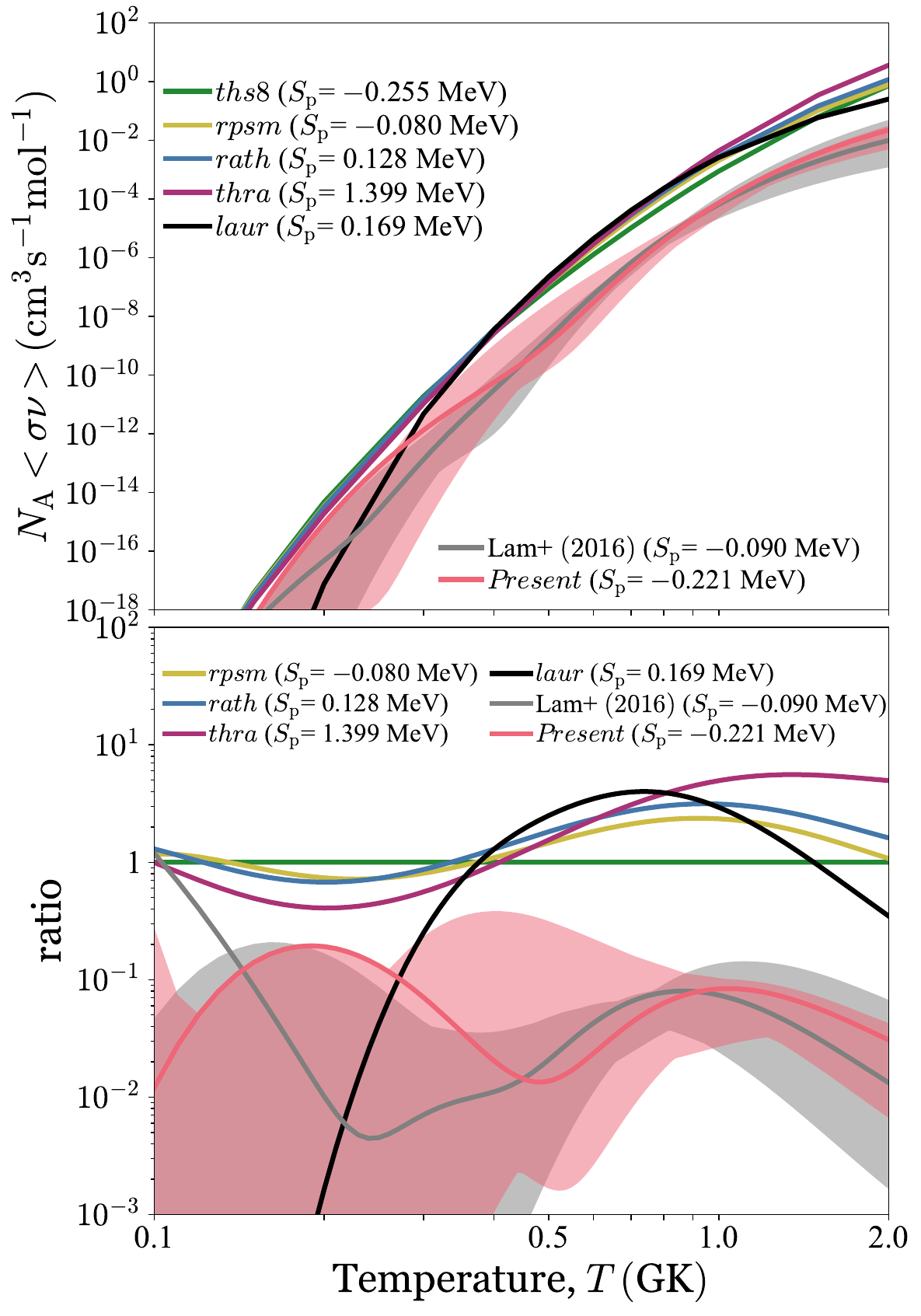}
\caption{\label{fig:rp_64Ge65As}The comparison of $^{64}$Ge(p,$\gamma$)$^{65}$As thermonuclear reaction rates. Top Panel: \citet{Lam2016}, \emph{ths}8, \emph{rath}, \emph{rpsm}, and \emph{thra} rates are the available rates compiled by \citet{Cyburt2010} and \emph{ths}8 is the recommended rate published in part of the JINA REACLIB~v2.2 release. 
Bottom Panel: the respective ratio of the reaction rates presented in the Top Panel to the recommended  \emph{ths}8 rate. The uncertainty of the present rate is indicated as red zone, whereas the uncertainty of the \citet{Lam2016} rate is shown as gray zone. 
}
\end{figure}


\renewcommand{\arraystretch}{0.85}
\LTcapwidth=\textwidth
\begin{table*}[tb]
\footnotesize
\caption{Properties of $^{64}$Ge of all dominant proton capture resonances for the present $^{63}$Ga(p,$\gamma$)$^{64}$Ge resonant rate calculation. The full list of resonances is presented in Tables~\ref{tab:ground} and \ref{tab:excited}\textcolor{black}{, which can be downloaded from the Zenodo repository~\citep{zenodo_Tables}.}}
\label{tab:properties_64Ge}
\begin{tabular*}{\linewidth}{@{\hspace{2mm}\extracolsep{\fill}}cllcccclll@{\hspace{2mm}}}
\toprule
\midrule
$J^{\pi}_i$  & $E_\mathrm{x}$ [MeV]\footnotemark[1] & $E_\mathrm{res}$ [MeV]\footnotemark[2]\textcolor{red}{$^,$}\footnotemark[3] & $C^2S_{7/2}$ & $C^2S_{3/2}$ & $C^2S_{5/2}$ &  $C^2S_{1/2}$ & $\Gamma_\gamma$ [eV]  & $\Gamma_\mathrm{p}$ [eV]  & $\omega\gamma$ [eV] \\
 &  &  & $(\mathtt{l}=3)$     & $(\mathtt{l}=1)$     & $(\mathtt{l}=3)$      &  $(\mathtt{l}=1)$      &  &  &  \\
\midrule
\multicolumn{10}{l}{Dominant resonances for proton captures on the ground state of $^{63}$Ga:}\\
$ 5^+_{  7} $ &$5.310$   & $0.253$   & $0.0004$ & $      $ & $      $ & $      $ & $1.731\times10^{-03}$ & $1.342\times10^{-16}$ & $1.845\times10^{-16}$ \\
$ 3^+_{ 15} $ &$5.356$   & $0.299$   & $0.0009$ & $0.0242$ & $0.0346$ & $      $ & $3.296\times10^{-02}$ & $2.698\times10^{-14}$ & $2.361\times10^{-14}$ \\
$ 1^+_{ 13} $ &$5.465$   & $0.408$   & $      $ & $0.0089$ & $0.0208$ & $0.0060$ & $6.819\times10^{-02}$ & $5.061\times10^{-11}$ & $1.898\times10^{-11}$ \\
$ 2^+_{ 28} $ &$5.622$   & $0.565$   & $0.0014$ & $0.0473$ & $0.0307$ & $0.0089$ & $2.114\times10^{-01}$ & $2.836\times10^{-07}$ & $1.772\times10^{-07}$ \\
$ 2^+_{ 33} $ &$5.855$   & $0.798$   & $0.0012$ & $0.0115$ & $0.0002$ & $0.0039$ & $1.386\times10^{-01}$ & $4.926\times10^{-05}$ & $3.078\times10^{-05}$ \\
$ 2^+_{ 36} $ &$5.946$   & $0.889$   & $0.0004$ & $0.0010$ & $0.0293$ & $0.0082$ & $2.903\times10^{-01}$ & $1.532\times10^{-04}$ & $9.570\times10^{-05}$ \\
$ 5^+_{ 14} $ &$5.992$   & $0.935$   & $0.0000$ & $      $ & $      $ & $      $ & $3.513\times10^{-03}$ & $1.842\times10^{-04}$ & $2.407\times10^{-04}$ \\
$ 3^+_{ 39} $ &$6.281$   & $1.224$   & $0.0003$ & $0.0066$ & $0.0087$ & $      $ & $1.668\times10^{-01}$ & $1.349\times10^{-02}$ & $1.092\times10^{-02}$ \\
$ 3^+_{ 42} $ &$6.430$   & $1.373$   & $0.0000$ & $0.0050$ & $0.0002$ & $      $ & $1.278\times10^{-01}$ & $4.630\times10^{-02}$ & $2.974\times10^{-02}$ \\
$ 1^+_{ 28} $ &$6.462$   & $1.330$   & $0.0000$ & $0.0001$ & $0.0008$ & $      $ & $5.775\times10^{-01}$ & $3.397\times10^{-04}$ & $8.488\times10^{-05}$ \\
$ 2^+_{ 50} $ &$6.462$   & $1.405$   & $0.0005$ & $0.0010$ & $0.0000$ & $0.0048$ & $2.438\times10^{-01}$ & $6.295\times10^{-02}$ & $3.127\times10^{-02}$ \\
$ 5^+_{ 24} $ &$6.537$   & $1.480$   & $0.0010$ & $      $ & $      $ & $      $ & $5.770\times10^{-02}$ & $4.031\times10^{-02}$ & $3.263\times10^{-02}$ \\
$ 2^+_{ 53} $ &$6.571$   & $1.514$   & $0.0001$ & $0.0036$ & $0.0003$ & $0.0004$ & $3.184\times10^{-01}$ & $1.277\times10^{-01}$ & $5.697\times10^{-02}$ \\
$ 1^+_{ 32} $ &$6.640$   & $1.583$   & $      $ & $0.0120$ & $0.0238$ & $0.0005$ & $9.387\times10^{-01}$ & $6.913\times10^{-01}$ & $1.493\times10^{-01}$ \\
$ 2^+_{ 55} $ &$6.652$   & $1.595$   & $0.0001$ & $0.0046$ & $0.0031$ & $0.0040$ & $5.224\times10^{-01}$ & $4.650\times10^{-01}$ & $1.538\times10^{-01}$ \\
$ 3^+_{ 51} $ &$6.700$   & $1.643$   & $0.0002$ & $0.0044$ & $0.0009$ & $      $ & $1.980\times10^{-01}$ & $3.556\times10^{-01}$ & $1.113\times10^{-01}$ \\
$ 3^+_{ 58} $ &$6.836$   & $1.779$   & $0.0000$ & $0.0030$ & $0.0003$ & $      $ & $4.454\times10^{-01}$ & $5.856\times10^{-01}$ & $2.214\times10^{-01}$ \\
\midrule
\multicolumn{10}{l}{Dominant resonances for proton captures on the first-excited state of $^{63}$Ga:}\\
$ 2^+_{ 24} $ &$5.382$   & $0.250$   & $0.0000$ & $0.0241$ & $0.0216$ & $0.0757$ & $5.570\times10^{-02}$ & $5.963\times10^{-16}$ & $2.485\times10^{-16}$ \\
$ 3^+_{ 19} $ &$5.520$   & $0.388$   & $0.0026$ & $0.0436$ & $0.0513$ & $0.0372$ & $2.563\times10^{-02}$ & $8.112\times10^{-11}$ & $4.732\times10^{-11}$ \\
$ 2^+_{ 31} $ &$5.691$   & $0.559$   & $0.0001$ & $0.0003$ & $0.0017$ & $0.0020$ & $4.817\times10^{-03}$ & $1.167\times10^{-08}$ & $4.862\times10^{-09}$ \\
$ 3^+_{ 23} $ &$5.702$   & $0.570$   & $0.0001$ & $0.0001$ & $0.0030$ & $0.0054$ & $5.508\times10^{-02}$ & $2.903\times10^{-08}$ & $1.693\times10^{-08}$ \\
$ 4^+_{ 23} $ &$5.726$   & $0.594$   & $0.0101$ & $0.0243$ & $0.0052$ & $      $ & $5.046\times10^{-02}$ & $2.836\times10^{-07}$ & $2.127\times10^{-07}$ \\
$ 6^+_{  9} $ &$5.732$   & $0.600$   & $0.0002$ & $      $ & $      $ & $      $ & $3.498\times10^{-03}$ & $1.638\times10^{-07}$ & $1.774\times10^{-07}$ \\
$ 6^+_{ 11} $ &$5.950$   & $0.818$   & $0.0002$ & $      $ & $      $ & $      $ & $3.419\times10^{-03}$ & $5.465\times10^{-05}$ & $5.827\times10^{-05}$ \\
$ 2^+_{ 38} $ &$5.995$   & $0.863$   & $0.0003$ & $0.0225$ & $0.0040$ & $0.0079$ & $8.541\times10^{-02}$ & $3.435\times10^{-04}$ & $1.426\times10^{-04}$ \\
$ 3^+_{ 39} $ &$6.281$   & $1.149$   & $0.0004$ & $0.0118$ & $0.0000$ & $0.0002$ & $1.668\times10^{-01}$ & $1.051\times10^{-02}$ & $5.767\times10^{-03}$ \\
$ 2^+_{ 46} $ &$6.335$   & $1.203$   & $0.0010$ & $0.0077$ & $0.0182$ & $0.0038$ & $2.617\times10^{-01}$ & $1.838\times10^{-02}$ & $7.156\times10^{-03}$ \\
$ 4^+_{ 39} $ &$6.363$   & $1.231$   & $0.0001$ & $0.0104$ & $0.0040$ & $      $ & $1.548\times10^{-01}$ & $2.258\times10^{-02}$ & $1.478\times10^{-02}$ \\
$ 4^+_{ 41} $ &$6.403$   & $1.271$   & $0.0031$ & $0.0191$ & $0.0177$ & $      $ & $3.415\times10^{-02}$ & $6.743\times10^{-02}$ & $1.700\times10^{-02}$ \\
$ 6^+_{ 17} $ &$6.446$   & $1.314$   & $0.0002$ & $      $ & $      $ & $      $ & $6.171\times10^{-02}$ & $5.724\times10^{-02}$ & $3.217\times10^{-02}$ \\
$ 3^+_{ 45} $ &$6.547$   & $1.415$   & $0.0010$ & $0.0021$ & $0.0048$ & $0.0013$ & $3.121\times10^{-01}$ & $4.513\times10^{-02}$ & $2.300\times10^{-02}$ \\
$ 3^+_{ 47} $ &$6.583$   & $1.451$   & $0.0001$ & $0.0038$ & $0.0114$ & $0.0003$ & $1.861\times10^{-01}$ & $6.654\times10^{-02}$ & $2.859\times10^{-02}$ \\
$ 2^+_{ 55} $ &$6.652$   & $1.520$   & $0.0001$ & $0.0076$ & $0.0003$ & $0.0077$ & $5.224\times10^{-01}$ & $4.736\times10^{-01}$ & $1.035\times10^{-01}$ \\
$ 3^+_{ 51} $ &$6.700$   & $1.568$   & $0.0012$ & $0.0008$ & $0.0121$ & $0.0032$ & $1.980\times10^{-01}$ & $1.722\times10^{-01}$ & $5.373\times10^{-02}$ \\
$ 3^+_{ 53} $ &$6.715$   & $1.583$   & $0.0004$ & $0.0056$ & $0.0017$ & $0.0007$ & $4.108\times10^{-01}$ & $3.404\times10^{-01}$ & $1.086\times10^{-01}$ \\
\bottomrule 
\end{tabular*}
\footnotetext[1]{Given by present large-scale shell model calculation.}
\footnotetext[2]{Calculated by $E_\mathrm{res}\!=\!E_\mathrm{x} - S_\mathrm{p}$ \textcolor{black}{for the proton captures on the ground state of $^{63}$Ga}, where \textcolor{black}{$S_\mathrm{p}\!=\!5.057\!\pm\!0.004$}~MeV (AME2020 \cite{AME2020}).}
\footnotetext[3]{\textcolor{black}{Calculated by $E_\mathrm{res}\!=\!E_\mathrm{x} - S_\mathrm{p} - E_i$ for the proton captures on the first excited state of $^{63}$Ga, where \textcolor{black}{$S_\mathrm{p}\!=\!5.057\!\pm\!0.004$}~MeV \citep{AME2020}, and $E_{i=5/2^-_1}\!=\!0.07510\!\pm\!0.00008$~MeV~\citep{NNDC_63Ga_2024}.}}
\end{table*}
\renewcommand{\arraystretch}{1.0}

\renewcommand{\arraystretch}{0.85}
\LTcapwidth=\textwidth
\begin{table*}[tb]
\vspace{-5mm}
\footnotesize
\caption{Properties of $^{65}$As of proton capture resonances for the present $^{64}$G\lowercase{e}(p,$\gamma$)$^{65}$A\lowercase{s} resonant rate calculation.}
\label{tab:properties_65As}
\begin{tabular*}{\linewidth}{@{\hspace{2mm}\extracolsep{\fill}}cllcccclll@{\hspace{2mm}}}
\toprule
\midrule
$J^{\pi}_i$ & & $E_\mathrm{x}$ [MeV] &  & $E_\mathrm{res}$ [MeV]\footnotemark[4] & $\mathtt{nlj}$ & $C^2S$ & $\Gamma_\gamma$ [eV]  & $\Gamma_\mathrm{p}$ [eV]  & $\omega\gamma$ [eV] \\
 & $E_\mathrm{x}^{\mathrm{Exp}}$\footnotemark[1]  & $E_\mathrm{x}^{\mathrm{Theo}}$\footnotemark[2]  & $E_\mathrm{x}$($^{65}$Ge)\footnotemark[3]      &  &   &   &  &  &  \\
\midrule
$ {3/2}^{\textcolor{red}{-}}_{1} $ &$0.000$    & $0.000$   & $0.000$ & $0.221$ & $2p_{3/2} $ & $0.196$ & $1.19\times10^{-34}$ & $3.87\times10^{-18}$ & $0.00$\\
$ {5/2}^{\textcolor{red}{-}}_{1} $ &$0.187(3)$ & $0.103$   & $0.111$ & $0.408$\footnotemark[5] & $1f_{5/2} $ & $0.533$ & $8.19\times10^{-17}$ & $3.28\times10^{-12}$ & $2.46\times10^{-16}$ \\
$ {5/2}^{\textcolor{red}{-}}_{2} $ &           & $0.501$   & $0.605$ & $0.722$ & $1f_{5/2} $ & $0.010$ & $3.76\times10^{-10}$ & $7.65\times10^{-9}$ & $1.13\times10^{-9}$ \\
$ {5/2}^{\textcolor{red}{-}}_{3} $ &           & $0.863$   &         & $1.084$ & $1f_{5/2} $ & $0.014$ & $1.64\times10^{-6}$  & $1.22\times10^{-5}$ & $4.89\times10^{-6}$ \\
$ {7/2}^{\textcolor{red}{-}}_{1} $ &           & $0.947$   & $0.890$ & $1.168$ & $1f_{7/2} $ & $0.013$ & $1.28\times10^{-5}$  & $7.52\times10^{-5}$ & $4.88\times10^{-5}$ \\
$ {7/2}^{\textcolor{red}{-}}_{2} $ &           & $1.070$   & $1.155$ & $1.291$ & $1f_{7/2} $ & $0.002$ & $8.50\times10^{-6}$  & $3.98\times10^{-5}$ & $3.27\times10^{-5}$ \\
\bottomrule
\end{tabular*}
\footnotetext[1]{Experimentally determined by~\citet{Obertelli2011}.}
\footnotetext[2]{Given by present large-scale shell model calculation.}
\footnotetext[3]{Compiled by~\citet{Browne2010}.}
\footnotetext[4]{Calculated by $E_\mathrm{res}\!=\!E_\mathrm{x} - S_\mathrm{p}$, where \textcolor{black}{$S_\mathrm{p}\!=\!-0.221\!\pm\!0.042$}~MeV~\citep{Zhou2023}.}
\footnotetext[5]{Experimental value of \textcolor{black}{$E_\mathrm{x}\!=\!0.187\!\pm\!0.003$}~MeV is used to calculate the $\Gamma_\gamma$ and $\Gamma_\mathrm{p}$ of this state.}
\end{table*}
\renewcommand{\arraystretch}{1.0}

The proton-capture spectroscopic factors, proton widths ($\Gamma_\mathrm{p}$), and gamma widths ($\Gamma_\gamma$) are obtained from the $^{63}$Ga and $^{64}$Ge nuclear wave functions based on the GXPF1A interaction. The nuclear spectroscopic information of dominant resonances are listed in Table~\ref{tab:properties_64Ge}, and the full list of nuclear spectroscopic information is presented in Tables~\ref{tab:ground} and \ref{tab:excited} in Appendix. The main contributing resonances of proton captures on the $3/2^-_\mathrm{g.s.}$ and $5/2^-_1$ states of $^{63}$Ga in the temperature region of XRB interest are plotted in the top and bottom panels of Fig.~\ref{fig:rp_63Ga_64Ge_contri}, respectively. The total present rate is shown as well. It is to be observed that the contribution from the resonance proton capture on the $5/2^-_1$ state of $^{63}$Ga is not negligible, and is up to about \textcolor{black}{a factor of 2-4} higher than the Hauser-Feshbach model rate (\emph{ths}8 or NON-SMOKER) at temperature, $T\!=\!1$-$2$~GK. The NON-SMOKER $^{63}$Ga(p,$\gamma$)$^{64}$Ge rate was released as the recommended rate in the JINA reaction library data set~\citep{Cyburt2010}, which was commonly used in recent XRB studies, e.g., the works of \citet{Zhou2023}, \citet{Lam2022a,Lam2022b}, \citet{Hu2021}, \citet{Randhawa2020}, \citet{Johnston2020}, and \citet{Goodwin2019}. 

For the $^{64}$Ge(p,$\gamma$)$^{65}$As thermonuclear reaction rate, the first-excited state of $^{64}$Ge is at \textcolor{black}{$E_\mathrm{x}\!=\!0.902\!\pm\!0.003$}~MeV, causing an exponentially low contribution to the resonant rate, and thus the proton capture on the thermally excited state can be neglected. Unlike the high level density of $^{64}$Ge, the energy levels of $^{65}$As at the respective Gamow energy are relatively sparse. We recalculate the resonance energies based on the newly determined $S_\mathrm{p}$($^{65}$As), and then use the nuclear spectroscopic information from Table~1 of \citet{Lam2016} to update the $^{64}$Ge(p,$\gamma$)$^{65}$As reaction rate. There are only six resonances of $E_\mathrm{res}\!=\!0.221$, $0.408$, $0.722$, $1.084$, $1.168$ and $1.291$~MeV dominating the total resonant rate in the temperature region of $0.2$-$2$~GK (Fig.~\ref{fig:rp_64Ge65As_contri}). The resonance properties of $^{65}$As for the proton capture on the ground-state of $^{64}$Ge are summarized in Table~\ref{tab:properties_65As}.

\vspace{-7mm}
\subsection{Direct-capture rates}
\label{sec:DC}

The direct proton capture rate on the target nucleus at state $i$ is given by~\citep{Schatz2005}
\begin{eqnarray}
\label{eq:DC}
N_\mathrm{A}\langle \sigma v \rangle_\mathrm{DC}^i = && 7.83 \times 10^9 \left (\frac{Z_\mathrm{T}}{\mu{T_9}^2}\right)^{1/3} \times S_\mathrm{DC}^i(E_0)\nonumber\\
&&\times  \;\; \mathrm{exp}\left (-4.249\left (\frac{Z_\mathrm{T}^2\mu}{T_9}\right)^{1/3} \right)\, 
\end{eqnarray}
\textcolor{black}{The reaction rate is expressed in units of $\mathrm{cm^3s^{-1}mol^{-1}}$.} 
Here $Z_\mathrm{T}$ is proton number of target nucleus, for $^{63}$Ga of the $^{63}$Ga(p,$\gamma$)$^{64}$Ge reaction, $Z_\mathrm{T}\!=\!31$ and for $^{64}$Ge of the $^{64}$Ge(p,$\gamma$)$^{65}$As reaction, $Z_\mathrm{T}\!=\!32$. $S_\mathrm{DC}^i(E_0)$ is the astrophysical $S$-factor at the Gamow energy, which can be expressed in terms of the $S$-factor at zero energy $S^i(0)$ and the dimensionless parameter $\tau$ = 4.2487$(Z_\mathrm{T}^2\mu/T_9)^{-1/3}$ as follows~\citep{Fowler1964,Rolfs1988}.
\vspace{-5mm}
\begin{eqnarray}
\label{eq:S-factor}
S_\mathrm{DC}^i(E_0) =  S^i(0) \left (1+\frac{5}{12\tau}\right) \,
\end{eqnarray}

For the present work, we used the \textsc{RadCap} code~\citep{Bertulani2003} to calculate the DC $S$-factor. The spin-orbit potential depth, $V_\mathrm{s.o.}\!=\!-10$~MeV, is taken from \citet{Huang2010}, and the rest of optical-potential parameters are $R_0\!=\!R_\mathrm{s.o.}\!=\!R_C\!=\!1.25\!\times\!(1\!+\!A_T)^{1/3}$~fm, and $a_0\!=\!a_\mathrm{s.o.}\!=\!0.65$~fm. The depth of the central potential, $V_0$, is varied to reproduce the bound-state energies. 

For the $^{63}$Ga(p,$\gamma$)$^{64}$Ge reaction, the $S^i(0)$ of the ground state and the first-excited state $E_{i=5/2^-}\!=\!0.075\!\pm\!0.00018$~MeV in $^{63}$Ga  are \textcolor{black}{$10.6$}~MeV$\cdot$b and \textcolor{black}{$0.1$}~MeV$\cdot$b, respectively. The contribution of the direct proton capture rate, $N_\mathrm{A}\langle \sigma v \rangle_\mathrm{DC}^i$ to the total $^{63}$Ga(p,$\gamma$)$^{64}$Ge reaction rate is only \textcolor{black}{$0.003$~\%}, exponentially lower than the contribution of the dominating resonance rates throughout the XRB temperature range from $0.3$ to $2$~GK. For the $^{64}$Ge(p,$\gamma$)$^{65}$As reaction, the $S^i(0)$ is \textcolor{black}{24.3}~MeV$\cdot$b. The DC reaction rate, which is supposedly dominant at the low temperature regime, \textcolor{black}{$0.09\!\lesssim\!T\mathrm{(GK)}\!\lesssim\!0.20$}~GK, contributes merely about \textcolor{black}{$1.91$~\%} to the total reaction rate. By taking into account the estimated upper limits of the DC contribution~\citep{He2014}, the dominant contributions to each total present $^{63}$Ga(p,$\gamma$)$^{64}$Ge and $^{64}$Ge(p,$\gamma$)$^{65}$As reaction rates are still the respective resonant rates. Therefore, the DC contributions for both $^{63}$Ga(p,$\gamma$)$^{64}$Ge and $^{64}$Ge(p,$\gamma$)$^{65}$As reactions are in fact negligible.

\section{Discussions}
\label{sec:discussion}

Figure~\ref{fig:rp_63Ga_64Ge} shows the comparison of the present $^{63}$Ga(p,$\gamma$)$^{64}$Ge rate with other available reaction rates compiled in JINA REACLIB v2.2 by \citet{Cyburt2010}. The Hauser-Feshbach statistical model rates, i.e., \emph{ths}8, \emph{rpsm}, \emph{rath}, and \emph{thra} \citep{Rauscher2000} are very close to one another from $0.2$ to $2.0$~GK, with a difference of merely less than \textcolor{black}{a factor of 2}, and these statistical-model rates are lower than the present rate up to about \textcolor{black}{one order of magnitude}. Note that the statistical-model rates may include unknown systematic errors because of the limited capability of Hauser-Feshbach model in estimating the level densities of nuclei near to the proton drip line. The contribution from the proton capture resonances on the first excited state of $^{63}$Ga enlarges the gap between the present and \emph{ths}8 rates, causing a difference of up to \textcolor{black}{a factor of 4} at the temperature region, $T\!=\!0.8$-$2$~GK, which covers typical maximum temperature of the accreted envelopes of the GS~1826$-$24 periodic bursts and SAX~J1808.4$-$3658 PRE bursts, see the comparison of the respective ratio in the bottom panel of Fig.~\ref{fig:rp_63Ga_64Ge}. 

The comparison of the present $^{64}$Ge(p,$\gamma$)$^{65}$As rate with other available reaction rates in the JINA REACLIB v2.2, i.e., \emph{ths}8, \emph{rpsm}, \emph{rath}, \emph{thra}, \emph{laur} \citep{Rauscher2000}, and \citet{Lam2016}, is presented in Fig.~\ref{fig:rp_64Ge65As}. See \citet{Lam2016} for more details of the comparison between shell-model and statistical-model reaction rates. In this work, the main contributing resonance energies presented in Table~1 of \citet{Lam2016} are updated using the new \textcolor{black}{$S_\mathrm{p}$($^{65}$As)~$\!=\!-221\!\pm\!42$}~keV~\citep{Zhou2023}. We find that the shifts of these resonance energies cause the enlargement of uncertainty at \textcolor{black}{$0.23\!\lesssim\!T\mathrm{(GK)}\!\lesssim\! 0.97$}, whereas the uncertainty at $1\!\lesssim\!T\mathrm{(GK)}\!\lesssim\!2$ is reduced up to about \textcolor{black}{one order of magnitude}, compared to the \citet{Lam2016} $^{64}$Ge(p,$\gamma$)$^{65}$As rate. Moreover, at \textcolor{black}{$0.15\!\lesssim\!T\mathrm{(GK)}\!\lesssim\! 0.45$}, the present $^{64}$Ge(p,$\gamma$)$^{65}$As rate is up to about \textcolor{black}{two orders of magnitude} higher than the \citet{Lam2016} rate due to the newly determined $S_\mathrm{p}$($^{65}$As), which is \textcolor{black}{131}~keV lower than the previous \textcolor{black}{$S_\mathrm{p}(^{65}\mathrm{As})\!=\!-90\!\pm\!85$}~keV. At the temperature region of $0.5\!\lesssim\!T\mathrm{(GK)}\!\lesssim\!1.6$, which is more sensitive to the GS~1826$-$24 clocked bursts and SAX~J1808.4$-$3658 PRE bursts, the difference between both rates is up to merely \textcolor{black}{a factor of 1.9}.

Tables~\ref{tab:rate_Ga63pg} and \ref{tab:rate_Ge64pg} give the present $^{63}$Ga(p,$\gamma$)$^{64}$Ge and $^{64}$Ge(p,$\gamma$)$^{65}$As reaction rates in the temperature-rate grid format, respectively. Tables~\ref{tab:parameter_Ga63pg} and \ref{tab:parameter_Ge64pg} provide the corresponding parameterizations. See the note of Table~\ref{tab:parameter_Ga63pg} for reconstructing the rates presented in Tables~\ref{tab:rate_Ga63pg} and \ref{tab:rate_Ge64pg}. 
The uncertainties of the present $^{63}$Ga(p,$\gamma$)$^{64}$Ge and $^{64}$Ge(p,$\gamma$)$^{65}$As reaction rates are a combination of the theoretical nuclear spectroscopic uncertainty, $200$~keV, and experimental uncertainties, e.g., proton thresholds. 
The folded uncertainties yield several orders of magnitude of the upper and lower limits in the resonant rates for both $^{63}$Ga(p,$\gamma$)$^{64}$Ge and $^{64}$Ge(p,$\gamma$)$^{65}$As reactions (Figs.~\ref{fig:rp_63Ga_64Ge} and \ref{fig:rp_64Ge65As}). Future precise measurements for the spectroscopic information of the low-lying structure of $^{65}$As would shed light on further reducing the uncertainty of the $^{64}$Ge(p,$\gamma$)$^{65}$As reaction rate.



\renewcommand{\arraystretch}{0.8}
\begin{table}
\footnotesize
\caption{\label{tab:rate_Ga63pg}Thermonuclear reaction rates of $^{63}$Ga(p,$\gamma$)$^{64}$Ge as a function of temperature corresponding to XRB, given in the centroid, lower, and upper limits.}
\begin{tabular*}{\linewidth}{@{\hspace{2mm}\extracolsep{\fill}}cccc@{\hspace{2mm}}}
\toprule
\midrule[0.25pt]
$T_{9}$ & centroid & lower limit & upper limit \\

\midrule[0.25pt]
$0.1$  &   $3.66\times10^{-22}$ &   $1.12\times10^{-22}$ & $2.66\times10^{-21}$ \\ 
$0.2$  &   $1.14\times10^{-14}$ &   $1.14\times10^{-14}$ & $3.67\times10^{-14}$ \\ 
$0.3$  &   $9.34\times10^{-11}$ &   $5.48\times10^{-11}$ & $1.26\times10^{-10}$ \\
$0.4$  &   $1.94\times10^{-08}$ &   $1.42\times10^{-08}$ & $2.45\times10^{-08}$ \\
$0.5$  &   $8.38\times10^{-07}$ &   $6.39\times10^{-07}$ & $1.08\times10^{-06}$ \\
$0.6$  &   $1.46\times10^{-05}$ &   $1.18\times10^{-05}$ & $1.85\times10^{-05}$ \\
$0.7$  &   $1.38\times10^{-04}$ &   $1.27\times10^{-04}$ & $1.70\times10^{-04}$ \\
$0.8$  &   $8.60\times10^{-04}$ &   $8.60\times10^{-04}$ & $1.01\times10^{-03}$ \\    
$0.9$  &   $3.99\times10^{-03}$ &   $3.99\times10^{-03}$ & $4.40\times10^{-03}$ \\    
$1.0$  &   $1.47\times10^{-02}$ &   $1.47\times10^{-02}$ & $1.67\times10^{-02}$ \\    
$1.1$  &   $4.53\times10^{-02}$ &   $4.47\times10^{-02}$ & $5.14\times10^{-02}$ \\    
$1.2$  &   $1.19\times10^{-01}$ &   $1.13\times10^{-01}$ & $1.34\times10^{-01}$ \\    
$1.3$  &   $2.78\times10^{-01}$ &   $2.54\times10^{-01}$ & $3.08\times10^{-01}$ \\    
$1.4$  &   $5.83\times10^{-01}$ &   $5.22\times10^{-01}$ & $6.35\times10^{-01}$ \\    
$1.5$  &   $1.12\times10^{+00}$ &   $9.89\times10^{-01}$ & $1.20\times10^{+00}$ \\    
$1.6$  &   $2.01\times10^{+00}$ &   $1.75\times10^{+00}$ & $2.12\times10^{+00}$ \\    
$1.7$  &   $3.37\times10^{+00}$ &   $2.94\times10^{+00}$ & $3.53\times10^{+00}$ \\    
$1.8$  &   $5.38\times10^{+00}$ &   $4.68\times10^{+00}$ & $5.58\times10^{+00}$ \\    
$1.9$  &   $8.21\times10^{+00}$ &   $7.15\times10^{+00}$ & $8.46\times10^{+00}$ \\    
$2.0$  &   $1.21\times10^{+01}$ &   $1.05\times10^{+01}$ & $1.23\times10^{+01}$ \\  
\bottomrule[1.0pt]
\end{tabular*}
\end{table}
\renewcommand{\arraystretch}{1.0}

\renewcommand{\arraystretch}{0.8}
\begin{table}
\footnotesize
\caption{\label{tab:rate_Ge64pg}Thermonuclear reaction rates of $^{64}$Ge(p,$\gamma$)$^{65}$As as a function of temperature corresponding to XRB, given in the centroid, lower, and upper limits.}
\begin{tabular*}{\linewidth}{@{\hspace{2mm}\extracolsep{\fill}}cccc@{\hspace{2mm}}}
\toprule
\midrule[0.25pt]
$T_{9}$ & centroid & lower limit & upper limit \\

\midrule[0.25pt]
$0.1$  &   $1.34\times10^{-25}$ &   $6.12\times10^{-32}$ & $3.12\times10^{-24}$\\ 
$0.2$  &   $9.06\times10^{-16}$ &   $1.52\times10^{-19}$ & $9.13\times10^{-16}$\\ 
$0.3$  &   $1.34\times10^{-12}$ &   $5.75\times10^{-16}$ & $4.41\times10^{-12}$\\
$0.4$  &   $5.65\times10^{-11}$ &   $2.89\times10^{-12}$ & $1.07\times10^{-09}$\\
$0.5$  &   $1.24\times10^{-09}$ &   $1.51\times10^{-10}$ & $2.86\times10^{-08}$\\
$0.6$  &   $3.01\times10^{-08}$ &   $3.37\times10^{-09}$ & $2.85\times10^{-07}$\\
$0.7$  &   $4.52\times10^{-07}$ &   $8.25\times10^{-08}$ & $1.71\times10^{-06}$\\
$0.8$  &   $3.75\times10^{-06}$ &   $1.02\times10^{-06}$ & $7.75\times10^{-06}$\\    
$0.9$  &   $1.97\times10^{-05}$ &   $5.94\times10^{-06}$ & $2.82\times10^{-05}$\\    
$1.0$  &   $7.38\times10^{-05}$ &   $2.37\times10^{-05}$ & $9.13\times10^{-05}$\\    
$1.1$  &   $2.16\times10^{-04}$ &   $7.75\times10^{-05}$ & $2.59\times10^{-04}$\\    
$1.2$  &   $5.25\times10^{-04}$ &   $2.14\times10^{-04}$ & $6.31\times10^{-04}$\\    
$1.3$  &   $1.11\times10^{-03}$ &   $4.31\times10^{-04}$ & $1.35\times10^{-03}$\\    
$1.4$  &   $2.08\times10^{-03}$ &   $7.20\times10^{-04}$ & $2.58\times10^{-03}$\\    
$1.5$  &   $3.58\times10^{-03}$ &   $1.12\times10^{-03}$ & $4.53\times10^{-03}$\\    
$1.6$  &   $5.72\times10^{-03}$ &   $1.63\times10^{-03}$ & $7.37\times10^{-03}$\\    
$1.7$  &   $8.61\times10^{-03}$ &   $2.26\times10^{-03}$ & $1.13\times10^{-02}$\\    
$1.8$  &   $1.23\times10^{-02}$ &   $3.02\times10^{-03}$ & $1.64\times10^{-02}$\\    
$1.9$  &   $1.69\times10^{-02}$ &   $3.89\times10^{-03}$ & $2.30\times10^{-02}$\\    
$2.0$  &   $2.25\times10^{-02}$ &   $4.87\times10^{-03}$ & $3.13\times10^{-02}$\\  
\bottomrule[1.0pt]
\end{tabular*}
\end{table}
\renewcommand{\arraystretch}{1.0}


\renewcommand{\arraystretch}{0.8}
\begin{table*}
\footnotesize
\caption{\label{tab:parameter_Ga63pg}Parameters of $^{63}$Ga(p,$\gamma$)$^{64}$Ge centroid reaction rate.}
\begin{tabular*}{\linewidth}{@{\hspace{2mm}\extracolsep{\fill}}cccccccc@{\hspace{2mm}}}
\toprule
\midrule[0.25pt]
 $i$ & $a_0$                   & $a_1$                 & $a_2$                     & $a_3$                   & $a_4$                   & $a_5$                   & $a_6$      \\
\hline
$1$& $3.836500\times10^{+1}$ & $1.611100\times10^{-2}$ & $-4.185740\times10^{+1}$ & $ 1.858040\times10^{+0}$ & $-2.918800\times10^{+0}$ & $ 3.543180\times10^{-1}$ & $-4.127667\times10^{-1}$ \\
\bottomrule[1.0pt]
\end{tabular*}
\footnotetext[1]{The present $^{63}$Ga(p,$\gamma$)$^{64}$Ge centroid reaction rate can be reproduced by substituting the $a_0$, ..., $a_6$ parameters listed here to the expression proposed by~\citet{Rauscher2000}, 
$N_\mathrm{A}\langle\sigma v\rangle\!=\!\sum \mathrm{exp}(a_0 + \frac{a_1}{T_9} + \frac{a_2}{T_9^{1/3}} + a_3 T_9^{1/3} + a_4 T_9 + a_5 T_9^{5/3} + a_6 \ln{T_9})$, with an accuracy quantity, \textcolor{black}{$\zeta\!=\!2.72\times10^{-2}$} and the fitting error is \textcolor{black}{$5.40$}~\% for $T\!=\!0.1$-$2$~GK.}
\end{table*}
\renewcommand{\arraystretch}{1.0}

\renewcommand{\arraystretch}{0.8}
\begin{table*}
\footnotesize
\caption{\label{tab:parameter_Ge64pg}Parameters of $^{64}$Ge(p,$\gamma$)$^{65}$As centroid reaction rate.}
\begin{tabular*}{\linewidth}{@{\hspace{2mm}\extracolsep{\fill}}crrrrrrr@{\hspace{2mm}}}
\toprule
\midrule[0.25pt]
 $i$ & $a_0$                   & $a_1$                 & $a_2$                     & $a_3$                   & $a_4$                   & $a_5$                   & $a_6$      \\
\hline

$1$& $-7.81560\times10^{+1}$ & $-1.42680\times10^{+1}$ & $1.22180\times10^{+1}$ & $8.19260\times10^{+1}$ &$-1.30180\times10^{+1}$ & $1.75100\times10^{+0}$ &$-1.56690\times10^{+1}$ \\
$2$& $-9.38210\times10^{+1}$ & $-9.72000\times10^{+0}$ & $1.51390\times10^{+1}$ & $6.12270\times10^{+1}$ & $2.03370\times10^{+1}$ &$-9.34500\times10^{+0}$ &$-2.47030\times10^{+1}$ \\
$3$& $-7.98140\times10^{+1}$ & $-4.25800\times10^{+0}$ & $1.97070\times10^{+1}$ & $5.23700\times10^{+1}$ &$-3.23470\times10^{+1}$ & $1.46760\times10^{+1}$ & $1.28000\times10^{+0}$ \\
$4$& $-5.79490\times10^{+1}$ & $-3.09100\times10^{+0}$ &$-4.00000\times10^{-2}$ & $5.66640\times10^{+1}$ &$-1.16780\times10^{+1}$ & $1.11100\times10^{+0}$ &$-1.48000\times10^{+0}$ \\
$5$& $-1.16837\times10^{+2}$ & $-5.44300\times10^{+0}$ & $6.35010\times10^{+1}$ & $4.83650\times10^{+1}$ & $8.27210\times10^{+1}$ &$-1.42859\times10^{+2}$ & $2.14940\times10^{+1}$ \\
\bottomrule[1.0pt]
\end{tabular*}
\footnotetext[1]{The running index $i$ is up to 5 for the present $^{64}$Ge(p,$\gamma$)$^{65}$As centroid reaction rate reproduced by the same method mentioned in Table~\ref{tab:parameter_Ga63pg}, with an accuracy quantity, \textcolor{black}{$\zeta\!=\!9.30\times10^{-3}$} and the fitting error is \textcolor{black}{$5.65$}~\% for $T\!=\!0.1$-$2$~GK.}
\end{table*}
\renewcommand{\arraystretch}{1.0}



\begin{figure}[t]
\begin{center}
\vspace{5mm}
\includegraphics[width=8.7cm, angle=0]{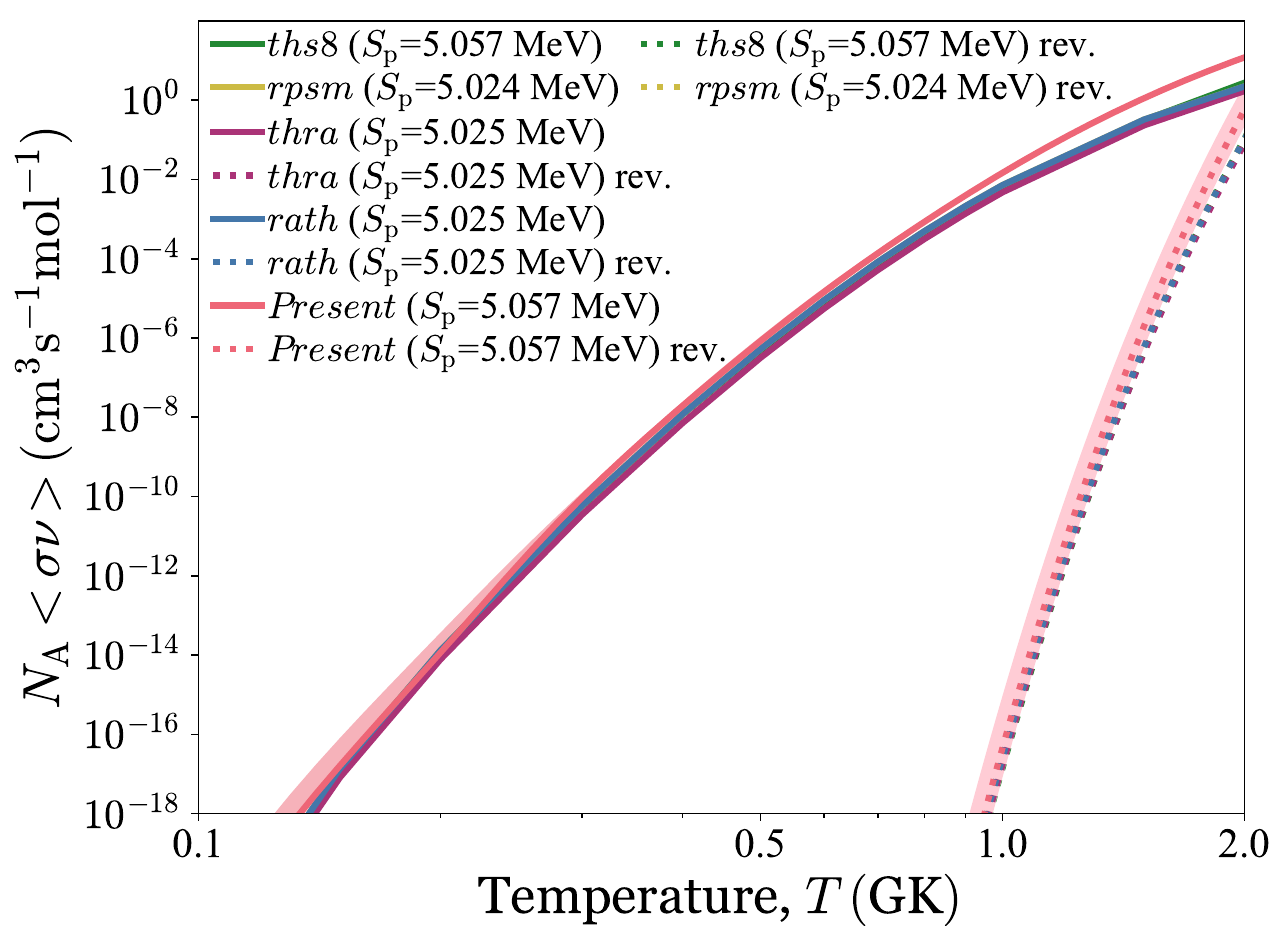}
\caption{\label{fig:63Ga_64Ge_reverse}{The forward and reverse thermonuclear reaction rates of $^{63}$Ga(p,$\gamma$)$^{64}$Ge in the temperature region of the XRB interest. The \emph{ths}8 forward and reverse rates are the recommended rate published in part of the JINA REACLIB~v2.2 release. The uncertainty of the present forward (reverse) rate is indicated as red (light red) zone.}}
\end{center}
\vspace{-5mm}
\end{figure}

\begin{figure}[t]
\begin{center}
\vspace{5mm}
\includegraphics[width=8.7cm, angle=0]{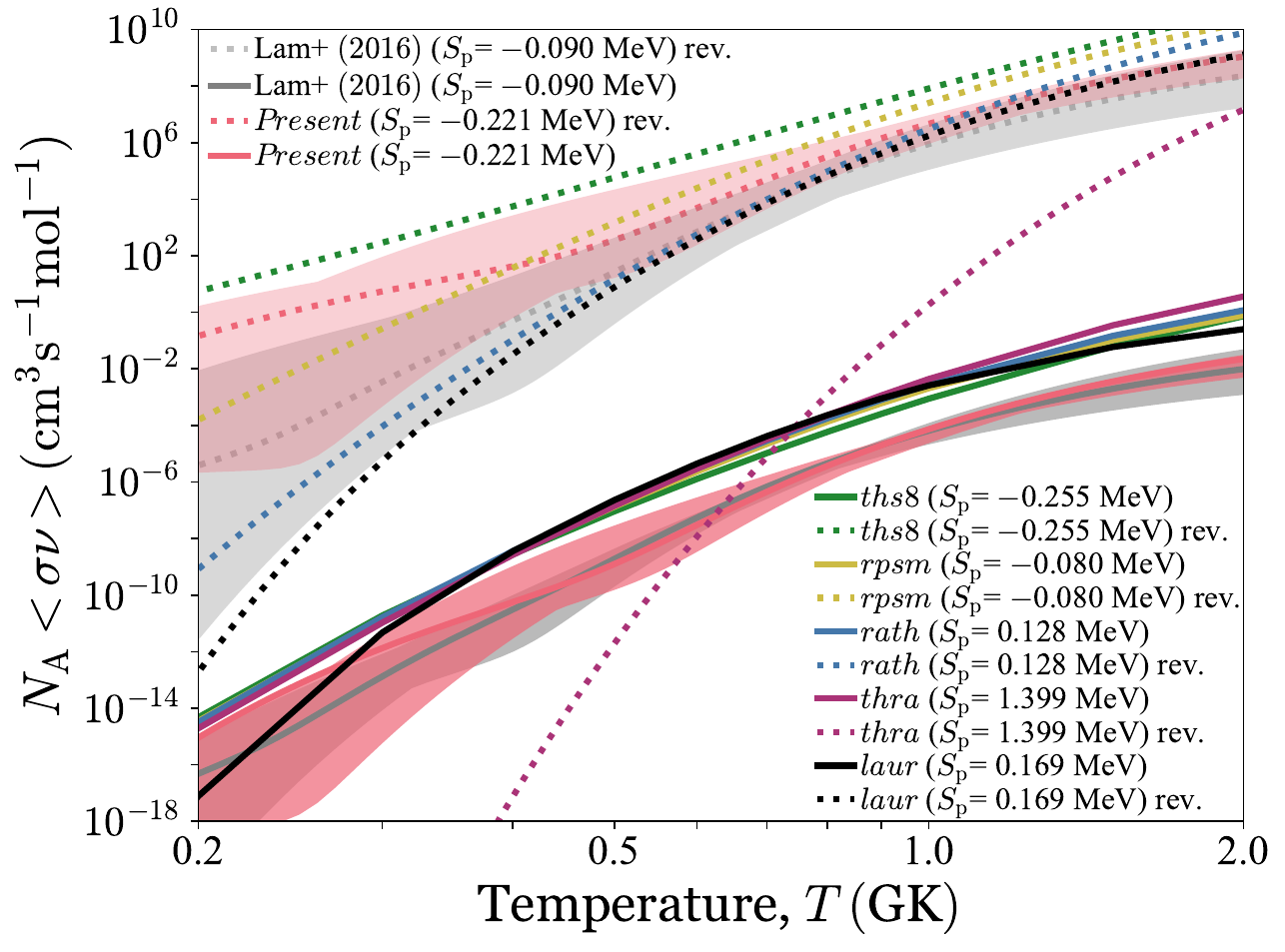}
\caption{\label{fig:64Ge_65As_reverse}{The forward and reverse thermonuclear reaction rates of $^{64}$Ge(p,$\gamma$)$^{65}$As in the temperature region of the XRB interest. The reverse rates are located at the upper region, $10^{-14}$ to $10^{2}$~cm$^3$s$^{-1}$mol$^{-1}$. The uncertainty of the present forward (reverse) rate is indicated as red (light red) zone, whereas the uncertainty of the \citet{Lam2016} forward (reverse) rate is shown as gray (light gray) zone.}}
\end{center}
\vspace{-5mm}
\end{figure}

\section{Summary and Prospective}
\label{sec:summary}

We have performed the large-scale shell-model calculations using the full $pf$-model space with the GXPF1A Hamiltonian to construct new $^{63}$Ga(p,$\gamma$)$^{64}$Ge and $^{64}$Ge(p,$\gamma$)$^{65}$As thermonuclear reaction rates (Figs.~\ref{fig:rp_63Ga_64Ge}, \ref{fig:rp_64Ge65As}, \ref{fig:63Ga_64Ge_reverse}, and \ref{fig:64Ge_65As_reverse}). The present work also estimates the uncertainty of these reaction rates, which is obtained from a combination of overall experimental and theoretical nuclear spectroscopic uncertainties (Figs.~\ref{fig:rp_63Ga_64Ge} and \ref{fig:rp_64Ge65As}). 
The construction of the $^{63}$Ga(p,$\gamma$)$^{64}$Ge reaction rate requires the nuclear spectroscopic information of resonance states in $^{64}$Ge above the newest proton-emission threshold of $5.057\pm0.004$~MeV \cite{AME2020}. The $^{63}$Ga(p,$\gamma$)$^{64}$Ge reaction rate is dominated by a large amount of resonances (Fig.~\ref{fig:level_density_64Ge}). The proton capture on the first excited state ($5/2^-_1$; $E_{5/2^-}\!=\!75.83\pm0.18$~keV) of $^{63}$Ga has also been taken into account as it is close to the ground state of $^{63}$Ga and enhances the resonant rate up to \textcolor{black}{a factor of $3$}, at $T\!=\!0.3$-$2.0$~GK (Fig.~\ref{fig:rp_63Ga_64Ge}), compared to the JINA REACLIB v2.2 recommended statistical-model $^{63}$Ga(p,$\gamma$)$^{64}$Ge reaction rate (\emph{ths}8). 
For deducing the new thermonuclear reaction rate of $^{64}$Ge(p,$\gamma$)$^{65}$As, the new proton separation energy of $^{65}$As ($-0.221\!\pm\!0.054$~MeV) from the newly measured $^{65}$As mass~\citep{Zhou2023} is implemented. 
The new $^{64}$Ge(p,$\gamma$)$^{65}$As rate is close to the \citet{Lam2016} rate from around $T\!=\!0.5$-$2.0$~GK, and is up to \textcolor{black}{two orders of magnitude} lower than the JINA REACLIB v2.2 recommended statistical-model rate (\emph{ths}8) at around $T\!=\!0.2$-$2.0$~GK (Fig.~\ref{fig:rp_64Ge65As}).


The new $^{63}$Ga(p,$\gamma$)$^{64}$Ge and $^{64}$Ge(p,$\gamma$)$^{65}$As thermonuclear reaction rates can be used for studying the thermo-hydrodynamics, extent of nucleosynthesis, and reproduction of available observables of type-I x-ray bursts.

As the prospective use of these reaction rates, we have implemented them to the state-of-the-art one-dimensional multi-zone thermo-hydrodynamic code, \textsc{Kepler} \citep{Weaver1978, Woosley2004, Heger2007}, and studied the influence of these new reaction rates on the burst light curves, recurrence times, and fluences of the GS~1826$-$24 clocked burster and of SAX~J1808.4$-$3658 photospheric radius expansion burster. These two models are constrained to reproducing the observed burst peaks, light-curve profiles, fluences, and recurrence times. We find that the impact of the newly measured proton thresholds and respective proton-capture reactions on the burst light-curve profile of the GS~1826$-$24 clocked burster is not as significant as claimed by \citet{Zhou2023}, whereas the influence on SAX~J1808.4$-$3658 photospheric radius expansion bursts is apparent. We will publish the forefront study of the impact of these upper and lower limits of the new $^{63}$Ga(p,$\gamma$)$^{64}$Ge and $^{64}$Ge(p,$\gamma$)$^{65}$As reaction rates on the burst light-curve profiles, recurrence times, and fluences of GS~1826$-$24 and SAX~J1808.4$-$3658 bursters elsewhere.

\begin{acknowledgments}
\textcolor{black}{We deeply appreciate the anonymous Referees for carefully reviewing our manuscript with useful suggestions and comments.} 
We are very grateful to Michael S. Smith (Oak Ridge National Laboratory, US) for using the Computational Infrastructure for Nuclear Astrophysics to parameterize the new reaction rates. We thank Noritaka Shimizu (University of Tsukuba, Japan) for suggestions in tuning the \textsc{KShell} code at the FDR5 cluster (Academia Sinica Grid-computing Centre) of Academia Sinica, Taiwan. 
This work was supported by the Strategic Priority Research Program of \textcolor{magenta}{Chinese Academy of Sciences} (CAS, Grant Nos. \textcolor{teal}{XDB34020204} and \textcolor{teal}{XDB34020100}), \textcolor{magenta}{National Key R\&D Program of China} (No. \textcolor{teal}{2021YFA1601500}), and \textcolor{magenta}{National Natural Science Foundation of China} (Nos. \textcolor{teal}{12375146}, \textcolor{teal}{11775277}, \textcolor{teal}{11961141004}). We deeply appreciate the computing resources, i.e., Distributed Cloud resources (FDR5 cluster), provided by the Institute of Physics and Academia Sinica Grid-computing Center of Academia Sinica (ASGC, Grant No. AS-CFII-112-103), Taiwan. 
Y.H.L. gratefully acknowledges the financial supports from the \textcolor{magenta}{Chinese Academy of Sciences} President's International Fellowship Initiative (No. \textcolor{teal}{2019FYM0002}), the graceful hospitality from Shigeru Kubono and Daisuke Suzuki (RIKEN, the University of Tokyo, Japan) for visiting RIKEN, and the support from Katsuhisa Nishio (Japan Atomic Energy Agency, Japan) for participating the \href{https://asrc.jaea.go.jp/soshiki/gr/HENS-gr/NAPS2024/}{ASRC Workshop} \citep{ASRC_Workshop}. A.H. is supported by the Australian Research Council Centre of Excellence for Gravitational Wave Discovery (OzGrav, No. CE170100004) and for All Sky Astrophysics in 3 Dimensions (ASTRO 3D, No. CE170100013), is also supported, in part, by the US National Science Foundation under Grant No. PHY-1430152 (JINA Center for the Evolution of the Elements, JINA-CEE). \textcolor{black}{Z.X.L. is supported by Talent Support Project of Huizhou (Innovation Team 2023, No. 230806198702053).} 
\textcolor{black}{H.Y. is supported by \textcolor{magenta}{JSPS KAKENHI} (No. \textcolor{teal}{18H01218}, No. \textcolor{teal}{19K03883}, and No. \textcolor{teal}{23H01181}) from the Ministry of Education, Culture, Sports, Science and Technology (MEXT) of Japan.}
\end{acknowledgments}

\appendix

\section{Appendixes}

Table~\ref{tab:ground} lists the properties of $^{64}$Ge for the ground-state proton capture on $^{63}$Ga in the present $^{63}$Ga(p,$\gamma$)$^{64}$Ge resonant rate calculation. 

Table~\ref{tab:excited} lists the properties of $^{64}$Ge for the first-excited-state proton capture on $^{63}$Ga in the present $^{63}$Ga(p,$\gamma$)$^{64}$Ge resonant rate calculation.

\renewcommand{\arraystretch}{0.85}
\LTcapwidth=\textwidth
\scriptsize
\footnotesize
\begin{longtable*}{@{\hspace{2mm}\extracolsep{\fill}}cllcccclll@{\hspace{2mm}}}
\label{tab:ground}\\
\caption{Properties of $^{64}$Ge for the ground-state proton capture \textcolor{black}{on $^{63}$Ga} in the present $^{63}$G\lowercase{a}($p$,$\gamma$)$^{64}$G\lowercase{e} resonant rate calculation. \textcolor{black}{An electronic version of this table can be downloaded from the Zenodo repository~\citep{zenodo_Tables}.}} \\
\toprule
\midrule[0.25pt]
$J^{\pi}_i$  & $E_\mathrm{x}$ [MeV]$^\mathrm{a}$ & $E_\mathrm{res}$ [MeV]$^\mathrm{b}$ & $C^2S_{7/2}$ & $C^2S_{3/2}$ & $C^2S_{5/2}$ &  $C^2S_{1/2}$ & $\Gamma_\gamma$ [eV]  & $\Gamma_\mathrm{p}$ [eV]  & $\omega\gamma$ [eV] \\
             &  &  & $(l=3)$      & $(l=1)$      & $(l=3)$      &  $(l=1)$      &  &  &  \\
\hline
\endfirsthead
\caption[]{(continued)} \\
\toprule
\midrule[0.25pt]
$J^{\pi}_i$  & $E_\mathrm{x}$ [MeV]$^\mathrm{a}$ & $E_\mathrm{res}$ [MeV]$^\mathrm{b}$ & $C^2S_{7/2}$ & $C^2S_{3/2}$ & $C^2S_{5/2}$ &  $C^2S_{1/2}$ & $\Gamma_\gamma$ [eV]  & $\Gamma_\mathrm{p}$ [eV]  & $\omega\gamma$ [eV] \\
             &  &  & $(l=3)$      & $(l=1)$      & $(l=3)$      &  $(l=1)$      &  &  &  \\
\hline \\
\endhead
$ 1^+_{  8} $ &$5.118$     & $0.061$     & $      $ & $0.0002$ & $0.0115$ & $0.0025$ & $3.434\times10^{-03}$ & $1.118\times10^{-44}$ & $4.192\times10^{-45}$ \\
$ 5^+_{  5} $ &$5.128$     & $0.071$     & $0.0002$ & $      $ & $      $ & $      $ & $3.555\times10^{-03}$ & $8.444\times10^{-39}$ & $1.161\times10^{-38}$ \\
$ 3^+_{ 11} $ &$5.168$     & $0.111$     & $0.0001$ & $0.0041$ & $0.0008$ & $      $ & $3.403\times10^{-03}$ & $1.305\times10^{-30}$ & $1.142\times10^{-30}$ \\
$ 5^+_{  6} $ &$5.184$     & $0.127$     & $0.0000$ & $      $ & $      $ & $      $ & $2.206\times10^{-03}$ & $3.331\times10^{-28}$ & $4.580\times10^{-28}$ \\
$ 2^+_{ 20} $ &$5.190$     & $0.133$     & $0.0005$ & $0.0026$ & $0.0206$ & $0.0001$ & $4.125\times10^{-02}$ & $2.497\times10^{-27}$ & $1.561\times10^{-27}$ \\
$ 3^+_{ 12} $ &$5.198$     & $0.141$     & $0.0008$ & $0.0614$ & $0.0098$ & $      $ & $5.958\times10^{-02}$ & $5.956\times10^{-25}$ & $5.211\times10^{-25}$ \\
$ 3^+_{ 13} $ &$5.205$     & $0.148$     & $0.0005$ & $0.0004$ & $0.0084$ & $      $ & $3.303\times10^{-03}$ & $3.135\times10^{-26}$ & $2.743\times10^{-26}$ \\
$ 2^+_{ 21} $ &$5.224$     & $0.167$     & $0.0003$ & $0.0006$ & $0.0002$ & $0.0009$ & $3.782\times10^{-03}$ & $1.089\times10^{-23}$ & $6.806\times10^{-24}$ \\
\hline
\end{longtable*}
\begin{minipage}{\textwidth}
\noindent
$^\mathrm{a}$ Given by present large-scale shell model calculation.\\
$^\mathrm{b}$ Calculated by $E_\mathrm{res}\!=\!E_\mathrm{x} - S_\mathrm{p}$, where \textcolor{black}{$S_\mathrm{p}\!=\!5.057\!\pm\!0.004$}~MeV (AME2020 \cite{AME2020}). \\
$*$ represents the dominant resonance states.\\
$\dag$ indicates the resonance states, which are presented with the same resonance energies in three decimal points, but as a matter of fact these states are not degenerate for the reason that the nuclear shell-model wave functions of the considered nuclei are written in the $M$-scheme representation, which are stored in huge matrices of double precision \citep{Shimizu2022}. The diagonalization of huge matrices are computed in double precision \citep{KShell}. 
\end{minipage}

\newpage
\newpage
\renewcommand{\arraystretch}{0.85}
\LTcapwidth=\textwidth
\scriptsize
\footnotesize
\begin{longtable*}{@{\hspace{2mm}\extracolsep{\fill}}cllcccclll@{\hspace{2mm}}}
\label{tab:excited}\\
\caption{Properties of $^{64}$Ge for the \textcolor{black}{first-excited-state proton capture on $^{63}$Ga} in the present $^{63}$G\lowercase{a}($p$,$\gamma$)$^{64}$G\lowercase{e} resonant rate calculation. \textcolor{black}{An electronic version of this table can be downloaded from the Zenodo repository~\citep{zenodo_Tables}.}} \\
\toprule
\midrule[0.25pt]
$J^{\pi}_i$  & $E_\mathrm{x}$ [MeV]$^\mathrm{a}$ & $E_\mathrm{res}$ [MeV]$^\mathrm{b}$ & $C^2S_{7/2}$ & $C^2S_{3/2}$ & $C^2S_{5/2}$ &  $C^2S_{1/2}$ & $\Gamma_\gamma$ [eV]  & $\Gamma_\mathrm{p}$ [eV]  & $\omega\gamma$ [eV] \\
             &  &  & $(l=3)$      & $(l=1)$      & $(l=3)$      &  $(l=1)$      &  &  &  \\
\hline
\endfirsthead
\caption[]{(continued)} \\
\toprule
\midrule[0.25pt]
$J^{\pi}_i$  & $E_\mathrm{x}$ [MeV]$^\mathrm{a}$ & $E_\mathrm{res}$ [MeV]$^\mathrm{b}$ & $C^2S_{7/2}$ & $C^2S_{3/2}$ & $C^2S_{5/2}$ &  $C^2S_{1/2}$ & $\Gamma_\gamma$ [eV]  & $\Gamma_\mathrm{p}$ [eV]  & $\omega\gamma$ [eV] \\
             &  &  & $(l=3)$      & $(l=1)$      & $(l=3)$      &  $(l=1)$      &  &  &  \\
\hline \\
\endhead
$ 3^+_{ 12} $ &$5.198$     & $0.066$     & $0.0003$ & $0.0080$ & $0.0002$ & $0.0262$ & $5.958\times10^{-02}$ & $1.637\times10^{-41}$ & $9.549\times10^{-42}$ \\
$ 3^+_{ 13} $ &$5.205$     & $0.073$     & $0.0007$ & $0.0009$ & $0.0123$ & $0.0001$ & $3.303\times10^{-03}$ & $1.827\times10^{-40}$ & $1.066\times10^{-40}$ \\
$ 2^+_{ 21} $ &$5.224$     & $0.092$     & $0.0000$ & $0.0002$ & $0.0065$ & $0.0000$ & $3.782\times10^{-03}$ & $9.194\times10^{-36}$ & $3.831\times10^{-36}$ \\
$ 2^+_{ 22} $ &$5.235$     & $0.103$     & $0.0018$ & $0.0000$ & $0.0002$ & $0.0176$ & $9.079\times10^{-02}$ & $1.687\times10^{-31}$ & $7.029\times10^{-32}$ \\
$ 1^+_{  9} $ &$5.242$     & $0.110$     & $0.0000$ & $0.0205$ & $0.0071$ & $      $ & $2.767\times10^{-01}$ & $4.264\times10^{-30}$ & $1.066\times10^{-30}$ \\
$ 4^+_{ 15} $ &$5.267$     & $0.135$     & $0.0003$ & $0.0063$ & $0.0003$ & $      $ & $4.712\times10^{-03}$ & $1.017\times10^{-26}$ & $7.628\times10^{-27}$ \\
$ 4^+_{ 16} $ &$5.273$     & $0.141$     & $0.0062$ & $0.0347$ & $0.1555$ & $      $ & $9.430\times10^{-02}$ & $3.441\times10^{-25}$ & $2.581\times10^{-25}$ \\
$ 1^+_{ 10} $ &$5.300$     & $0.168$     & $0.0001$ & $0.0076$ & $0.0051$ & $      $ & $1.110\times10^{-02}$ & $6.766\times10^{-23}$ & $1.691\times10^{-23}$ \\
\hline
\end{longtable*}
\begin{minipage}{\textwidth}
\noindent
$^\mathrm{a}$ Given by present large-scale shell model calculation.\\
$^\mathrm{b}$ Calculated by $E_\mathrm{res}\!=\!E_\mathrm{x} - S_\mathrm{p} - E_i$, where \textcolor{black}{$S_\mathrm{p}\!=\!5.057\!\pm\!0.004$}~MeV \citep{AME2020}, and \textcolor{black}{$E_{i=5/2^-_1}\!=\!0.07510\!\pm\!0.00008$~MeV~\citep{NNDC_63Ga_2024}}. \\
$*$ represents the dominant resonance states.\\
$\dag$ indicates the resonance states, which are presented with the same resonance energies in three decimal points, but as a matter of fact these states are not degenerate for the reason that the nuclear shell-model wave functions of the considered nuclei are written in the $M$-scheme representation, which are stored in huge matrices of double precision \citep{Shimizu2022}. The diagonalization of huge matrices are computed in double precision \citep{KShell}. 
\end{minipage}



\bibliography{XRB_PRC_01}

\end{document}